\documentclass[preprintnumbers,nofootinbib,
superscriptaddress,amsmath,floatfix,prd]{revtex4}

\usepackage{amssymb}  
\usepackage{graphicx,subfigure}
\usepackage{exscale}
\usepackage{textcomp}
\usepackage{enumerate}

\usepackage{color}

\usepackage[normalem]{ulem}  

\newcommand{\non}{\nonumber\\}

\newcommand{\be}{\begin{equation}}
\newcommand{\ee}{\end{equation}}
\newcommand{\bea}{\begin{eqnarray}}
\newcommand{\eea}{\end{eqnarray}}
\newcommand{\ba}[1]{\begin{array}{#1}}
\newcommand{\ea}{\end{array}}

\newcommand{\Tr}{{\rm Tr}}

\begin{document}

\preprint{\hfill {\small {ICTS-USTC-15-09}}}

\title{From holography towards real-world nuclear matter}

\author{Si-wen Li}
\email{cloudk@mail.ustc.edu.cn}
\affiliation{Interdisciplinary Center for Theoretical Study and Department of Modern Physics, University of Science and Technology of China, Hefei 230026, China}

\author{Andreas Schmitt}
\email{aschmitt@hep.itp.tuwien.ac.at}
\affiliation{Institut f\"{u}r Theoretische Physik, Technische Universit\"{a}t Wien, 1040 Vienna, Austria}

\author{Qun Wang}
\email{qunwang@ustc.edu.cn}
\affiliation{Interdisciplinary Center for Theoretical Study and Department of Modern Physics, University of Science and Technology of China, Hefei 230026, China}

\date{16 July 2015}

\begin{abstract}

Quantum chromodynamics is notoriously difficult to solve at nonzero baryon density, and most models or effective theories of dense quark or nuclear matter 
are restricted to a particular density regime and/or a particular form of matter. Here we study dense (and mostly cold) matter within the holographic Sakai-Sugimoto model, aiming at 
a strong-coupling framework in the wide density range between nuclear saturation density and ultra-high quark matter densities. 
The model contains only three parameters, and we ask 
whether it fulfills two basic requirements of real-world cold and dense matter, 
a first-order onset of nuclear matter and a chiral phase transition at high density to quark matter. 
Such a model would be extremely useful for astrophysical applications because it would provide a single equation of state for all densities relevant in a compact star. 
Our calculations are based on two approximations for baryonic matter, firstly 
an instanton gas and secondly a 
homogeneous ansatz for the non-abelian gauge fields on the flavor branes of the model. While the instanton gas shows chiral restoration at high densities but 
an unrealistic second-order baryon onset, the homogeneous ansatz behaves exactly the other way around. 
Our study thus provides all ingredients that are necessary for a more realistic model and allows for systematic improvements of the applied approximations.

\end{abstract}

\maketitle

\section{Introduction}

\subsection{Motivation}

Quantum chromodynamics (QCD) at nonzero baryon densities and zero (or very small) temperatures is very difficult, and progress has mostly been made with the help of various effective theories and 
phenomenological models. What we know for certain is that there is a first-order transition from the vacuum to nuclear matter at nuclear saturation density $n_0\simeq 0.16\,{\rm fm}^{-3}$ 
and that 
QCD becomes weakly interacting at ultra-high densities due to asymptotic freedom \cite{Gross:1973id,Politzer:1973fx}, wherefore sufficiently dense matter is a weakly coupled gas of quarks. Hence there must be chiral and 
deconfinement phase transitions -- presumably at moderate densities well in the strong-coupling regime -- from nuclear to quark matter\footnote{Although
most models predict a first-order phase transition for small temperatures, it is conceivable that there is a smooth crossover \cite{Schafer:1998ef,Hatsuda:2006ps,Schmitt:2010pf}
from nuclear matter to color-flavor locked (CFL) quark matter \cite{Alford:1998mk,Alford:2007xm}. In this paper, we shall ignore color-superconducting phases 
such as the CFL phase, although it is an interesting question whether it can be implemented in the Sakai-Sugimoto model, possibly following existing attempts within holography 
\cite{Basu:2011yg}. Also, we shall work in the chiral limit, such that the transition between chirally broken baryonic matter and chirally restored quark matter must be a phase 
transition in the strict sense.}.    
All current rigorous studies are restricted to a particular density regime, for example, ``ordinary'' nuclear physics at and slightly above saturation density, lattice studies 
for baryon chemical potentials smaller than or at most of the order of the temperature, and perturbative 
studies at extremely large densities. But even giving up on some rigor, it is very difficult to come up with a reasonable model that describes QCD matter over a wide density regime.
For example, the quark-meson model, see for instance Refs.\  \cite{Schaefer:2007pw,Schaefer:2009ui}, and the Nambu-Jona Lasinio (NJL) model 
\cite{Nambu:1961tp,Nambu:1961fr}, see for instance 
Refs.\ \cite{Fukushima:2003fw,Buballa:2003qv,Ruester:2005jc}, and variants thereof can be very useful to 
get some insight into quark matter phases and possibly the chiral and 
deconfinement phase transitions, but they usually do not include nuclear matter, although it is possible \cite{Alkofer:1994ph}. 
On the other hand, nucleon-meson models, such as in Refs.\ \cite{Walecka:1974qa,Boguta:1977xi,Gallas:2011qp,Floerchinger:2012xd,Drews:2013hha}, 
are based on the properties of nuclear matter at saturation and may describe moderately dense nuclear matter realistically, but  
give a poor description, if at all, of chirally restored quark matter. Therefore, attempts to study strongly coupled dense matter from the onset of baryons all the way to 
quark matter at high densities are mostly based on patching together various models and, not surprisingly, depend on many parameters \cite{Berges:1998ha,Benic:2015pia}. 
Having a single model at hand -- and be it only a very distant relative of the fundamental theory -- 
would be very interesting not only from a theoretical point of view, but also for the applications in the context of compact stars. Compact stars have a density profile, and thus 
knowledge about the strongly coupled physics from nuclear saturation density possibly up to quark matter densities in the center of the star is required \cite{Schmitt:2010pn}. 

\subsection{Model}

The gauge-gravity correspondence \cite{Maldacena:1997re,Gubser:1998bc,Witten:1998qj} provides a powerful tool to study strongly coupled physics. 
Although the existence of a gravity dual of QCD is suggested by general principles, it is currently unknown and probably out of reach in the near future, 
and most studies have focused on theories that are different, sometimes very different, from QCD. The most prominent example is ${\cal N}=4$ super Yang-Mills theory 
where $N_f$ quark flavors  are introduced in the gravity dual by $N_f$ D7-branes in the background given by $N_c$ D3-branes; for studies in this setup aiming at the QCD phase diagram see for instance
Ref.\ \cite{Evans:2011eu} and, including baryonic matter, Refs.\  \cite{Gwak:2012ht,Evans:2012cx}.
The Sakai-Sugimoto model \cite{Witten:1998zw,Sakai:2004cn,Sakai:2005yt} is the holographic model that currently comes closest to QCD. In this paper we ask 
the question whether it can be employed in the dense matter context 
we have just outlined. Like the D3/D7 system, the Sakai-Sugimoto model is a ``top-down'' approach, i.e., it is rigorously based on an underlying type-IIA string theory, and the dual field theory is known. 
Namely, in a certain limit, it is dual to large-$N_c$ QCD. Here, we apply several approximations and extrapolations, giving up  some of this rigor in favor of  feasibility and of potentially coming 
closer to real-world $N_c=3$ QCD at large baryon chemical potentials: firstly, as in most previous applications of the model, 
we work in the classical gravity approximation, which is a good approximation for very large values of the 't Hooft coupling $\lambda$ 
(although we shall extrapolate some of our results to small $\lambda$),  
while large-$N_c$ QCD is the dual field theory only in the opposite, inaccessible, limit of small $\lambda$. Secondly, our main results are obtained in the 
``deconfined geometry'', which has a richer phase structure concerning chiral symmetry.  While the ``confined geometry'' can be connected to the confined phase 
of the dual field theory, this is less clear for the deconfined geometry \cite{Mandal:2011ws,Mandal:2011uq,Rebhan:2014rxa}. We thus do not a priori know the (phases of) the field
theory we are actually working in, especially regarding confinement/deconfinement (chiral symmetry, on the other hand, and its spontaneous breaking are implemented unambiguously). 
Thirdly, as in most 
related studies, we shall work in the probe brane limit, i.e., we shall not include the backreaction of the flavor branes on the background geometry (for studies going beyond this quenched
approximation, see Refs.\ \cite{Burrington:2007qd,Bigazzi:2014qsa}). And, we shall make use of a certain, not uniquely defined, prescription for the non-abelian 
Dirac-Born-Infeld (DBI) action to all orders in the string tension. Therefore, at very large densities, where effects of backreaction as well as higher-order terms in the 
string tension may become important, our results must be considered as an extrapolation from the low-density regime, where the approach is rigorous. 

\subsection{Goal and relation to previous works}

One important benefit of the given holographic approach is that there is a well defined way to implement baryons. 
Following the general idea of a ``baryon vertex'' in AdS/CFT \cite{Witten:1998xy,Gross:1998gk}, baryons in the Sakai-Sugimoto model are implemented as D4-branes, which have $N_c$ 
string endpoints, wrapped on the 4-sphere of the background geometry. Equivalently, they can be considered as instantons of the gauge field theory on the connected 
D8- and $\overline{\rm D8}$-branes \cite{Sakai:2004cn,Hong:2007kx,Hata:2007mb,Hong:2007ay}. 
Baryon properties have been computed  with the help of the Belavin-Polyakov-Schwarz-Tyupkin (BPST) instanton solution \cite{1975PhLB...59...85B}, 
using instantons in flat space as a trial function for the equations of motion of the curved geometry \cite{Hata:2007mb}, although improvements to this approach might be 
necessary \cite{Cherman:2009gb,Cherman:2011ve,Bolognesi:2013nja,Rozali:2013fna,Bolognesi:2013jba}. 
Here we are not interested in single baryons in vacuum, but in baryonic matter. We shall restrict ourselves to homogeneous matter; 
for studies of inhomogeneities and instanton crystals in the confined geometry 
of the Sakai-Sugimoto model, see Refs.\ \cite{Kaplunovsky:2012gb,deBoer:2012ij,Kaplunovsky:2015zsa}. 
A simple and very useful approximation for homogeneous baryonic matter makes use of pointlike instantons \cite{Bergman:2007wp} and has been 
employed in both confined and deconfined 
geometries. This approximation shows an unrealistic second-order phase transition from vacuum to baryonic matter (in both geometries), and, also unrealistically, 
low-temperature baryonic matter remains the preferred phase for arbitrarily large chemical potentials. In the confined geometry, this is obvious from the geometry of the model in the 
probe brane approximation and a consequence of the large-$N_c$ limit. (Although it might be possible to see signatures of chiral restoration also in the confined geometry 
\cite{Bolognesi:2014dja}, in particular by going beyond the probe brane approximation \cite{Bigazzi:2014qsa}.)
In the deconfined geometry, where chiral restoration at large chemical potentials is a priori possible, 
it turns out that the free energy 
of the baryonic phase asymptotically approaches that of the chirally restored phase, but always remains lower \cite{Preis:2011sp}. The main question we address in this paper is whether 
going beyond the pointlike approximation of baryons in the deconfined geometry can remedy one or both of these unphysical properties. It has been argued that, at 
least in the confined geometry, finite-size instantons 
indeed give rise to a first-order baryon onset \cite{Ghoroku:2012am,Ghoroku:2013gja}. The first part of our calculations is based on this finite-size instanton approach, 
but we shall argue that a fully dynamical calculation does {\it not} show a first-order onset, see Sec.\ \ref{sec:2nd} and in particular footnote \ref{footGhoroku}. 
The second part of our calculations makes use of a simple homogeneous ansatz instead of the instanton solution, which, for the confined geometry, 
has been introduced in Ref.\ \cite{Rozali:2007rx}.

\subsection{Outline of the paper}

We start with presenting the general setup in Sec.\ \ref{sec:setup}, where we introduce the gauge field action on the flavor D8-branes. The next two sections deal with our two different 
approaches: in Sec.\ \ref{sec:inst} we discuss the instanton gas and Sec.\ \ref{sec:hom} is devoted to the homogeneous ansatz. In the case of the instanton gas, we 
discuss the expansion of the Dirac-Born-Infeld (DBI) action for small non-abelian field strengths and the relation to the pointlike approximation in Sec.\ \ref{sec:instantons}. Our actual 
calculation is then performed using the full DBI action, and we derive the necessary equations and explain the numerical evaluation in Secs.\ \ref{sec:allorders} and \ref{sec:numerical}.
These sections are rather technical and readers only interested in the result may jump to Sec.\ \ref{sec:2nd} where we present our results (including the 
results for the confined geometry). Sec.\ \ref{sec:hom} is structured similarly: we first explain our approach and the calculation in Sec.\ \ref{sec:ancalc}, before we present the 
numerical results in Sec.\ \ref{sec:1st}. The main results are Figs.\ \ref{figdeconf1} and \ref{figphasesinst} for the instanton gas and Figs.\ \ref{figdeconf2} and \ref{figphaseshom} for the homogeneous ansatz. Finally, we give our conclusions in Sec.\ \ref{sec:conclusions}.

\section{Setup}
\label{sec:setup}

We start by setting the stage without going into details of or reviewing the Sakai-Sugimoto model since there are numerous works in the literature 
where this has been done, see for instance Refs.\ \cite{Bigazzi:2014qsa,Kaplunovsky:2015zsa,Peeters:2007ab,Rebhan:2008ur,Gubser:2009md,Kim:2012zzg,Rebhan:2014rxa}. 
In the following we focus on the features of the model that are relevant for us and write down the action we are using.
 
The Sakai-Sugimoto model incorporates a deconfinement phase transition by allowing for two different background geometries, and realizes a chiral phase transition 
by two different embeddings of the flavor branes in these background geometries.  In the original version of the model, the flavor D8- and 
$\overline{\rm D8}$-branes, corresponding to left- and right-handed fermions, are maximally separated at the boundary $U=\infty$, where $U$ is the holographic coordinate. 
Denoting this asymptotic separation by $L$, maximal separation means $L=\pi/M_{\rm KK}$, where the Kaluza-Klein mass $M_{\rm KK}$ is the inverse radius of the compactified 
extra dimension $X_4$ of the model, $X_4\equiv X_4 + 2\pi/M_{\rm KK}$. In this version, the chiral and deconfinement phase transitions coincide, i.e.,  
in the confined geometry the D8- and $\overline{\rm D8}$-branes are always connected, which corresponds to the chirally broken phase, and in the deconfined geometry 
they are always disconnected, indicating an unbroken chiral symmetry. Both phase transitions occur at a critical temperature $T_c=M_{\rm KK}/(2\pi)$ and do
not depend on the baryon chemical potential $\mu$
if the backreaction of the flavor branes on the background geometry is neglected. The reason is that the chemical potential is introduced via the gauge fields on the flavor branes 
and thus -- within this probe brane approximation -- cannot affect the background geometry and therefore has no effect on either deconfinement or chiral phase transitions.   

In this paper, we are mostly interested in non-antipodal asymptotic separations where $L\ll \pi/M_{\rm KK}$. This version of the model can be considered as the ``decompactified'' limit,
because it is achieved by either choosing $L$ very small compared to a fixed radius $M_{\rm KK}^{-1}$ or by keeping $L$ fixed and choosing a very large radius, i.e., 
by decompactifying the extra dimension. Of course, the dual field theory is only truly four-dimensional in the opposite limit, i.e., for a very {\it small} radius
of the extra dimension, and effects of the fifth dimension can be expected to become important in the decompactified limit. 
Since a large radius corresponds to a very small Kaluza-Klein mass, the critical temperature for deconfinement
goes to zero in the decompactified limit. Thus, when we work in the deconfined geometry, we shall ignore that below $T_c$ the confined geometry is preferred. 
Besides the Kaluza-Klein mass $M_{\rm KK}$ and the asymptotic separation of the flavor branes $L$, the only other free parameter of the model is the 't Hooft coupling $\lambda$.

For our purpose, the decompactified limit is most interesting because here the chiral phase transition is no longer locked to the deconfinement phase transition and
depends on $T$ {\it and} $\mu$. The resulting phase diagram in the $T$-$\mu$ plane in the absence of baryons \cite{Horigome:2006xu} (and including a background magnetic field
\cite{Preis:2010cq,Preis:2012fh}) shows striking similarities to that of a Nambu-Jona-Lasinio (NJL) model, 
in accordance with Ref.\ \cite{Antonyan:2006vw,Davis:2007ka}, where the connection to the NJL model was first pointed out. 
Two differences to the NJL model
are the current quark masses, which are very naturally included in NJL but very difficult to implement in Sakai-Sugimoto (although some attempts have been made 
\cite{Bergman:2007pm,Dhar:2007bz,Hashimoto:2008sr,Brunner:2015yha}), and the order of the chiral phase transition which, in the 
chiral limit, can be second or first order in NJL but is always first order in Sakai-Sugimoto. 

The asymptotic separation of the flavor branes $L$ therefore extrapolates between the original version of the Sakai-Sugimoto model, where the 
dual theory is large-$N_c$ QCD (at least in the inaccessible limit of small $\lambda$), and the decompactified limit, which has an NJL-like dual. 
Since dense 
matter at large $N_c$ is very different from that at $N_c=3$ \cite{Shuster:1999tn,McLerran:2007qj}, we are less interested in the more rigorous large-$N_c$ limit and focus our 
attention on the NJL-like 
limit, even though some important physics related to the gluon dynamics are thrown out. One may for instance ask what it really means to include baryons in a limit of 
the model where there is no confinement. We shall not discuss this question in detail and simply observe that instantons and thus baryon number 
can be introduced in the deconfined geometry completely analogously to the confined geometry.

The two scenarios in which we study baryonic matter are 
\begin{itemize}

\item {\it Confined geometry with antipodal separation, $L=\pi/M_{\rm KK}$ (free parameters: $\lambda$, $M_{\rm KK}$).} This scenario is  
used as a reference and preparation for the more complicated calculation in the deconfined geometry. The calculation is simpler than that of the deconfined geometry because the 
embedding of the connected branes does not have to be determined dynamically. All equations and derivations are deferred to the appendix; in the main part we only present 
the results, which are also useful as a comparison to the existing literature. The results are independent of temperature, and thus the only relevant thermodynamic variable is the 
baryon chemical potential.

\item {\it Deconfined geometry (free parameters: $L$, $\lambda$, $M_{\rm KK}$).} Our main results are obtained in this geometry and all derivations in the main text refer to this case.
The results depend on temperature and baryon chemical potential. Although we have in mind the decompactified limit when we discuss this geometry for all temperatures,
we do not make use of $L\ll \pi/M_{\rm KK}$ in the actual calculation and simply treat $L$ as a free parameter.

\end{itemize}

Our calculation starts from the gauge field action on the D8- and $\overline{\rm D8}$-branes, which consists of a Dirac-Born-Infeld (DBI) and a Chern-Simons (CS) contribution,
\be
S = S_{\rm DBI} + S_{\rm CS} \, .
\ee
We first discuss the DBI term, which has the form
\be \label{SDBI}
S_{\rm DBI} = 2T_8 V_4\int_0^{1/T} d\tau \int d^3X \int_{U_c}^\infty dU \, e^{-\Phi}\sqrt{{\rm det}(g+2\pi\alpha' {\cal F})} \, , 
\ee
where $\alpha'=\ell_s^2$ is the string tension with the string length $\ell_s$, $T_8 = 1/[(2\pi)^8\ell_s^9]$ is the D8-brane tension, and 
$e^{\Phi}= g_s(U/R)^{3/4}$ is the dilaton with the string coupling $g_s$ and the curvature radius of the background geometry $R$. We have already performed the 
trivial integration over the 4-sphere, resulting in the volume factor $V_4=8\pi^2/3$. The remaining integral is taken over Euclidean time $\tau$, 
3-dimensional position space, and the holographic coordinate $U$, from the tip of the connected
D8- and $\overline{\rm D8}$-branes $U=U_c$ to the holographic boundary $U=\infty$. This integral is taken over one half of the connected branes, hence the 
prefactor 2 to account for both halves. The induced metric on the flavor branes $g$ in the deconfined geometry is given by 
\begin{subequations}
\bea
ds_{\rm D8}^2 = \left(\frac{U}{R}\right)^{3/2}(f_Td\tau^2+\delta_{ij}dX^idX^j)+\left(\frac{R}{U}\right)^{3/2}\left\{\left[\frac{1}{f_T}+\left(\frac{U}{R}\right)^3
(\partial_UX_4)^2\right] dU^2 +U^2d\Omega_4^2\right\} \, ,
\eea
\end{subequations} 
where $i=1,2,3$, and $d\Omega_4^2$ is the metric of the 4-sphere. The function $X_4(U)$ describes the embedding of the flavor branes in the background geometry, and
\be
f_T \equiv  1-\frac{U_T^3}{U^3} \, , 
\ee
where $U_T$ is the location of the tip of the cigar-shaped $\tau$-$U$ subspace, related to temperature and curvature radius via
\be
T = \frac{3}{4\pi}\frac{U_T^{1/2}}{R^{3/2}} \, . 
\ee
The following relations between the parameters of the model will be useful,
\be \label{constants}
R^3 = \pi g_s N_c\ell_s^3 \,, \qquad \lambda = \frac{2R^{3}M_{\rm KK}}{\ell_s^2}  \, , 
\ee
where $N_c$ is the number of colors, i.e., the gauge group of the dual field theory is $SU(N_c)$.

We are considering two flavors, i.e., a $U(2)$ gauge theory in the bulk, whose local symmetry group corresponds to a global symmetry in the dual field theory. 
The field strength tensor is decomposed into a $U(1)$ and an $SU(2)$ part,
\be
{\cal F}_{\mu\nu} = \hat{F}_{\mu\nu} + F_{\mu\nu} \, , \qquad F_{\mu\nu}= F_{\mu\nu}^a \sigma_a \, ,
\ee
where $\mu,\nu=0,1,2,3,U$, and the Pauli matrices $\sigma_a$ ($a=1,2,3$). We shall use the same notation for the gauge fields, i.e., $\hat{A}_\mu$ and $A_\mu$ 
for abelian and non-abelian 
components, respectively. The physics we are interested in requires us to introduce a baryon chemical potential and baryons. 
The quark chemical potential (related to the baryon chemical potential by a factor $N_c$) is introduced as the boundary value of $\hat{A}_0$, 
while the baryons are introduced in the $SU(2)$ part via the field strengths $F_{ij}$, $F_{iU}$. Thus our only non-vanishing field strengths will be
$\hat{F}_{0U}=-\partial_U\hat{A_0}$, $F_{ij}$, $F_{iU}$. We shall set $\hat{F}_{i0}=\partial_i\hat{A_0}=0$ from the beginning because we are only interested in homogeneous matter, i.e., 
$\hat{A}_0$ will only depend on the holographic coordinate $U$, not on $\vec{X}$.

Within this ansatz, the CS contribution is
\be \label{SCS}
S_{\rm CS} = \frac{N_c}{4\pi^2}\int_0^{1/T} d\tau \int d^3X \int_{U_c}^\infty dU\, \hat{A}_0 \Tr[F_{ij}F_{kU}]\epsilon_{ijk} \, . 
\ee

\section{Baryonic matter from an instanton gas}
\label{sec:inst}

\subsection{Expansion in non-abelian field strengths}
\label{sec:instantons}

The DBI action as written in Eq.\ (\ref{SDBI}) does not define how to treat the case of non-abelian gauge fields. A general 
form, keeping all orders of the non-abelian gauge fields, is not known. In expansions for small $F$ (more precisely, for small string tension $\alpha'$), usually the symmetrized trace is used for all terms of ${\cal O}(F^4)$ 
and higher, which however is known to be incomplete starting from ${\cal O}(F^6)$ \cite{Sevrin:2001ha}. For our instanton gas approximation, we discuss two different approaches: 
in the current section, we expand the DBI action for small non-abelian field strengths up to quadratic order, while keeping the abelian field strength to all orders, 
following Ref.\ \cite{Ghoroku:2012am}. Later, in Sec.\ \ref{sec:allorders}, we shall employ an ``abelianized''  prescription that keeps abelian {\it and} non-abelian 
 field strengths to all orders and that we use for our numerical calculation (for our purposes, it is actually simpler not to expand the square root). 

For the expansion to second order in the non-abelian field strengths we need to replace the square root of the DBI action (\ref{SDBI}) as follows,   
\bea \label{expand}
\sqrt{{\rm det}(g+2\pi\alpha'{\cal F})}&\to&\sqrt{{\rm det}(g+2\pi\alpha'\hat{F})}\left\{1-\frac{(2\pi\alpha')^2}{4}
\Tr\left[(g+2\pi\alpha'\hat{F})^{-1} F\right]^2\right\}  + {\cal O}(F^4)\, , 
\eea
with 
\begin{subequations}\label{exp1}
\bea 
\sqrt{{\rm det}(g+2\pi\alpha'\hat{F})} &=& U^4\left(\frac{R}{U}\right)^{3/4}\sqrt{1+u^3 f_T x_4'^2-\hat{a}_0'^2} \, , \\[2ex] 
\Tr\left[(g+2\pi\alpha'\hat{F})^{-1} F\right]^2 &=& -\left(\frac{R}{U}\right)^3\Tr[F_{ij}^2]-\frac{2f_T\Tr[F_{iU}^2]}{1+u^3f_Tx_4'^2-\hat{a}_0'^2} \, , \label{exp12}
\eea
\end{subequations}
where the trace on the left-hand side of Eq.\ (\ref{exp12}) is taken over $8+1$-dimensional 
metric space and 2-dimensional flavor space, while the trace on the right-hand side is only taken over flavor space. We have 
replaced $\hat{A}_0\to i\hat{A}_0$, which is necessary since we work in Euclidean space-time, and the prime denotes derivative with respect to $u$. 
Moreover, we have introduced the dimensionless quantities $u$, $x_4$, $\hat{a}_0$, defined in Table \ref{table0} together with all other dimensionless 
quantities used in this paper\footnote{There is a slight difference 
in convention compared to Refs.\ \cite{Bergman:2007wp,Preis:2011sp}, which otherwise use the same notation: in that references, for example 
$\hat{a}_0 = 2\pi\alpha'/R\,\hat{A}_0$. We have included additional powers of the dimensionless factor $M_{\rm KK} R$ because then all observable quantities 
(chemical potential, baryon density, temperature) are directly obtained 
from their dimensionless counterparts by choosing values solely of $\lambda$ and $M_{\rm KK}$ (and not also values for $R$ and $\alpha'$, which have no direct interpretation in the 
dual field theory).}.

Inserting these results into the DBI action (\ref{SDBI}), we obtain 
\be \label{SDBI2}
S_{\rm DBI} = {\cal N} \int_0^{1/T} d\tau \int d^3X  \int_{u_c}^\infty du \, u^{5/2} \sqrt{1+u^3 f_T x_4'^2-\hat{a}_0'^2} \left[1+\frac{R^3(2\pi\alpha')^2}{4U^3}\Tr[F_{ij}^2]
+\frac{f_T}{2}\frac{(2\pi\alpha')^2\Tr[F_{iU}^2]}{1+u^3f_Tx_4'^2-\hat{a}_0'^2}\right] \, , 
\ee
where we have abbreviated 
\be \label{N}
{\cal N} \equiv \frac{2T_8 V_4}{g_s} R^5 (M_{\rm KK} R)^7 = \frac{N_c}{6\pi^2}\frac{R^2(M_{\rm KK} R)^7}{(2\pi\alpha')^3} \, .
\ee

\begin{table*}[t]
\begin{tabular}{|c|c|c|c|c|c|} 
\hline
\rule[-1.5ex]{0em}{6ex} 
 $\;x_4, \ell\;$ &$\;u,u_c,u_T,u_{\rm KK},z,\rho\;$  & $t$ & $q(z)$ & $\hat{a}_0,\mu,h$ & $n_I$ \\[2ex] \hline
\rule[-1.5ex]{0em}{6ex} 
$\;M_{\rm KK}\;$ & $\displaystyle{\frac{1}{R(M_{\rm KK} R)^2}}$ & $\;\;\displaystyle{\frac{1}{M_{\rm KK}}}\;\;$
& $\;R(M_{\rm KK} R)^2 \;$ & 
$\;\displaystyle{\frac{2\pi\alpha'}{R(M_{\rm KK} R)^2} = \frac{4\pi}{\lambda M_{\rm KK}}}\;$ & 
$\;\displaystyle{\frac{N_c}{\cal N}\frac{R(M_{\rm KK} R)^2}{2\pi\alpha'}=\frac{96\pi^4}{\lambda^2 M_{\rm KK}^3}}\;$  \\[2ex] \hline
\end{tabular}
\caption{Relation between the dimensionless quantities (first row) and their dimensionful counterparts, for example $x_4 = M_{\rm KK}X_4, t=T/M_{\rm KK}$, etc. 
Generally, we use capital letters for dimensionful and small letters for dimensionless quantities. Exceptions are $\rho$ (here we use only one symbol for both quantities, which 
does not cause confusion since both quantities appear always well separated from each other), $q(z)$ [where the dimensionful version is denoted by $q(Z)$], 
$\mu$ (for which we never introduce a dimensionful version), and $n_I$ (where the dimensionful version is $N_I/V$, $N_I$ being the instanton number, $V$ the 3-volume, and $n_I$ the dimensionless instanton number {\it density}).
}
\label{table0}
\end{table*}

Instead of solving the full equations of motion for abelian and non-abelian fields, we shall for simplicity employ an ansatz for the non-abelian field strengths that is based 
on the BPST instanton solution. We then will have to minimize the free energy with respect to the parameters that are introduced by this
ansatz (the number density of instantons and their width), together with solving the equations of motion for the abelian field $\hat{a}_0$ and the embedding function 
$x_4$. The instanton solution is best introduced in a coordinate that extends continuously over both halves of the connected flavor branes. This new coordinate 
$Z$ is defined via 
\be \label{Z}
U = (U_c^3+ U_c Z^2)^{1/3} \, , \qquad \frac{\partial U}{\partial Z} = \frac{2U_c^{1/2}\sqrt{f_c}}{3U^{1/2}} \, , 
\ee 
where
\be
f_c\equiv 1- \frac{U_c^3}{U^3} \, .
\ee
While $U\in [U_c,\infty]$, we have $Z\in[-\infty,\infty]$, where $Z=0$ corresponds to the tip of the connected branes. 
The instanton solution is well known for the flat-space Yang-Mills (YM) action. As explained
in appendix \ref{app:conf}, see Eq.\ (\ref{FijFiZ}), within this solution the traces of the field strengths become
\bea \label{F2s}
\Tr[F_{iZ}^2] &=&  \frac{6}{\gamma^2}\frac{4(\rho/\gamma)^4}{[\xi^2+(\rho/\gamma)^2]^4} \, , \qquad \Tr[F_{ij}^2] = 12\, \frac{4(\rho/\gamma)^4}{[\xi^2+(\rho/\gamma)^2]^4} 
\, , \qquad \Tr[F_{ij}F_{kZ}]\epsilon_{ijk} = - \frac{12}{\gamma}\frac{4(\rho/\gamma)^4}{[\xi^2+(\rho/\gamma)^2]^4} \, ,  
\eea
where we have used $\Tr[\sigma_a\sigma_a]=6$, abbreviated $\xi^2\equiv (\vec{X}-\vec{X}_0)^2+[(Z-Z_0)/\gamma]^2$, and   
\be
\gamma = \frac{3U_c^{3/2}}{2R^{3/2}} = \frac{3u_c^{3/2}}{2}(M_{\rm KK}R)^3 \, .
\ee
The instanton is located at the point $\vec{X}_0$ in position space and the point 
$Z_0$ in the bulk and its width is characterized by $\rho$.
The factor $\gamma$ was introduced in the BPST solution (\ref{AzAiinst}): 
while the instanton width in the holographic coordinate is $\rho$, it is $\rho/\gamma$ in position space. In the confined geometry with maximally separated flavor branes, 
$\gamma$ is just a constant, see Eq.\ (\ref{rho0}), and one can choose units in which $\gamma=1$. Therefore, the simplest form of the instanton used as a trial function in the 
Sakai-Sugimoto model is $SO(4)$ symmetric. We have generalized this instanton to the deconfined geometry in the most natural way, by simply replacing 
$U_{\rm KK}$ in $\gamma$ with its analogue in the deconfined geometry $U_c$. Since $U_c$ is not constant, we cannot work in units where $\gamma=1$. It has been argued that 
corrections beyond the limit of infinitely large 't Hooft coupling $\lambda$ lead to anisotropic instantons which break $SO(4)$ and which might be more realistic \cite{Rozali:2013fna}. 
For simplicity, we will not allow for these nontrivial configurations which go beyond the BPST ansatz, but it is instructive to keep in mind the role of $\gamma$ 
as a parameter for making the instantons anisotropic in a very simple way. 

In this paper, we are not interested in a single baryon, but in homogeneous baryonic matter. We thus consider a non-interacting 
gas of $N_I$ instantons
\cite{Ghoroku:2012am} located at positions $(\vec{X}_{0n},Z_{0n})$, $n=1,\ldots, N_I$. We shall assume that all instantons sit at the same point in the bulk, at the tip of the 
flavor branes, $Z_{n0}=0$ for all $n$. This is similar to the approximation from Ref.\ \cite{Bergman:2007wp}, where pointlike baryons were placed at the tip. Our approximation 
goes beyond the pointlike scenario, but does not determine the instanton distribution in the bulk dynamically. Moreover, we shall not solve the full $\vec{X}$-dependent 
equations of motion, but rather approximate the instanton distribution by its spatial average, such that the locations $\vec{X}_{0n}$ drop out. Therefore, we replace
\be \label{gas}
\frac{4(\rho/\gamma)^4}{[\xi^2+(\rho/\gamma)^2]^4} \to \frac{1}{V}\sum_{n=1}^{N_I}\int d^3 X \frac{4(\rho/\gamma)^4}{[(\vec{X}-\vec{X}_{0n})^2 + (Z/\gamma)^2 + (\rho/\gamma)^2]^4} 
= \frac{2\pi^2}{3} \gamma q(Z) \frac{N_I}{V} \,,  
\ee
where $V$ is the 3-volume, and we have defined the normalized function  
\be \label{qZ}
q(Z) \equiv \frac{3\rho^4}{4(Z^2+\rho^2)^{5/2}} \, , \qquad 
\int_{-\infty}^{\infty} dZ\,q(Z) = 1 \, .
\ee
Baryon number is generated by the $F^2$ term that couples to the abelian gauge field $\hat{A}_0$ in the CS term, because the abelian part of the gauge group $U(1)$ corresponds
to the global group at the boundary that is associated with baryon number conservation (by choosing the same chemical potential as a boundary value for $\hat{A}_0$ at both boundaries, 
$Z=+\infty$ and $Z=-\infty$, we ensure that we include a baryon chemical potential, not an axial chemical potential). We can thus write the baryon number as \cite{Sakai:2004cn}
\be \label{NNI}
-\frac{1}{8\pi^2}\int d^3X \int_{-\infty}^\infty dZ\, \Tr[F_{ij}F_{kZ}]\epsilon_{ijk} = N_I \, ,  
\ee
where we have inserted Eqs.\ (\ref{F2s}) and (\ref{gas}), i.e., the baryon number is identical to the number of instantons in our instanton gas. 
We can now write the CS part of the action as
\be \label{CS1}
S_{\rm CS} = -N_c \frac{V}{T}\frac{N_I}{V} \int_{-\infty}^\infty dZ\,\hat{A}_0 q(Z)  = -{\cal N} \frac{V}{T} n_I \int_{-\infty}^\infty dz\,\hat{a}_0 q(z)\, , 
\ee
with the dimensionless baryon number density $n_I$ and the dimensionless function $q(z)$ given in Table \ref{table0} [the function $q(z)$ depends on the 
dimensionless baryon width $\rho$ which, for notational convenience, we shall continue to denote with the same symbol as the dimensionful baryon width].
With the help of Eqs.\ (\ref{F2s}) and (\ref{gas}), we write all traces over the field strengths in terms of the baryon density, 
\bea \label{Fus} 
\frac{\Tr[F_{iU}^2]}{(M_{\rm KK}R)^3} \;\to\; \frac{1}{3\gamma}\frac{\partial z}{\partial u} \frac{n_Iq(u)}{(2\pi\alpha')^2} \, , \qquad 
\frac{\Tr[F_{ij}^2]}{(M_{\rm KK}R)^3} \;\to\;  \frac{2\gamma}{3}\frac{\partial u}{\partial z} \frac{n_Iq(u)}{(2\pi\alpha')^2} \, , \qquad 
\frac{\Tr[F_{ij}F_{kU}]\epsilon_{ijk}}{(M_{\rm KK}R)^3} \;\to\; - \frac{2}{3} \frac{n_Iq(u)}{(2\pi\alpha')^2} \,, 
\eea
where we have introduced
\be
q(u) \equiv 2\frac{\partial z}{\partial u} q(z) = \frac{9u^{1/2}}{4\sqrt{f_c}}\frac{(\rho^2u_c)^2}{(u^3-u_c^3+\rho^2u_c)^{5/2}} \, , \qquad \int_{u_c}^{\infty} du\,q(u) = 1 .  
\ee
We have included a factor 2 in the definition of $q(u)$ for convenience to ensure that $q(u)$ is normalized to one with respect to integration 
over one half of the connected flavor branes. 
The DBI action (\ref{SDBI2}), with the $\Tr[F^2]$ terms from Eq.\ (\ref{Fus}), together with the CS term (\ref{CS1}), yields the action 
\be
S = {\cal N}\frac{V}{T} \int_{u_c}^\infty du\, {\cal L} \, , 
\ee
with the Lagrangian 
\be \label{Lexp}
{\cal L}= {\cal L}_0
+n_Iq(u)\left[\frac{u\sqrt{f_T}}{6}\left(\frac{u_c^2 \sqrt{f_c}\sqrt{1+u^3f_T x_4'^2-\hat{a}_0'^2}}{u^2\sqrt{f_T}}+\frac{u^2\sqrt{f_T}}{u_c^2\sqrt{f_c}\sqrt{1+u^3f_T x_4'^2-\hat{a}_0'^2}}
\right)-\hat{a}_0\right] \, ,
\ee
where 
\be \label{L0}
{\cal L}_0 \equiv u^{5/2}\sqrt{1+u^3f_T x_4'^2-\hat{a}_0'^2}
\ee
is the DBI Lagrangian in the absence of instantons.

Let us discuss the relation of this Lagrangian to the approximation of pointlike instantons. In that case, the instanton profile is replaced by a delta function $\delta(u-u_c)$. 
The function $q(u)$ is the generalization of that delta function. The Lagrangian for pointlike instantons is \cite{Bergman:2007wp}
\be \label{Lpoint}
{\cal L}_{\rm pointlike} = {\cal L}_0 + n_I\left[\frac{u}{3}\sqrt{f_T(u)} - \hat{a}_0(u)\right]\delta(u-u_c) \, .
\ee
The first term is the same as in our Lagrangian (\ref{Lexp}), as well as the last term with $\delta(u-u_c)\to q(u)$. The second term is obtained from the action of $N_I$ D4-branes 
that are wrapped on the 4-sphere at $u=u_c$,
\be \label{SD4}
S_{\rm D4} = N_I T_4 \int d\Omega_4 d\tau e^{-\Phi} \sqrt{g} = {\cal N}\frac{V}{T}n_I\frac{u_c}{3}\sqrt{f_T(u_c)} \, , 
\ee
with the D4-brane tension $T_4=1/[(2\pi)^4\ell_s^5]$. The energy of the D4-branes $E_{\rm D4}=TS_{\rm D4}$ can then be interpreted as the mass of $N_I$ baryons. 
With the help of Table \ref{table0} we can write this energy as
\be \label{ED4}
E_{\rm D4} = N_I M_{\rm KK} \frac{\lambda N_c }{4\pi}\frac{u_c}{3}\sqrt{f_T(u_c)} \, .
\ee
We recover the baryon mass in the confined geometry with maximally separated flavor branes by setting $u_c\to u_{\rm KK} = 4/9$ and dropping the temperature-dependent factor. 
This yields $\lambda N_c M_{\rm KK}/(27\pi)$ as the mass of a single baryon to leading order in $\lambda$, in accordance with Refs.\ \cite{Sakai:2004cn,Hata:2007mb}.
In the deconfined geometry, $u_c\sqrt{f_T(u_c)}/3$ depends on chemical potential and temperature, giving rise to a medium-dependent baryon mass. 
We shall come back to this interpretation when we compare $u_c$ in our calculation with the result 
from the pointlike scenario, see Fig.\ \ref{figdeconf1}. 

\begin{figure} [t]
\begin{center}
\includegraphics[width=0.5\textwidth]{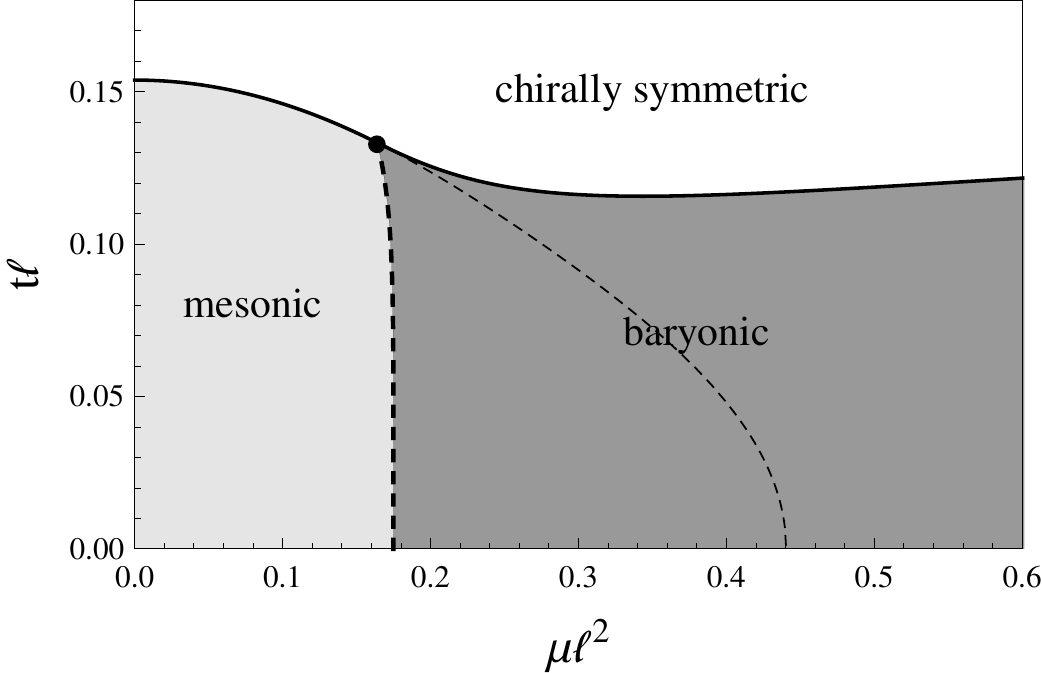}
\caption{Phase diagram in the plane of temperature and chemical potential in the deconfined geometry with 
pointlike baryons (first computed in Ref.\ \cite{Bergman:2007wp}). The baryon onset is a second-order phase transition (thick dashed line), while the chiral phase transition is 
first order (solid line). The thin dashed line would be the chiral phase transition if baryons were ignored (first computed in Ref.\ \cite{Horigome:2006xu}).  
At small temperatures, chiral symmetry remains broken for all values of the chemical potential.  We have rescaled temperature and chemical potential with the asymptotic separation of the flavor branes $\ell$. In all plots we use dimensionless scales, see Table \ref{table0}; we do not fit the parameters of the model since in this paper we are only interested in qualitative features of the phase diagram, and before the basic features of real-world dense matter are not reproduced by the model, any fit would be meaningless. The critical value 
for $t\ell$ below which the confined geometry is preferred can be made arbitrarily small in the decompactified limit, and thus the deconfinement transition is not seen here.}
\label{figpointlike}
\end{center}
\end{figure}

We show the phase diagram in the plane of temperature and chemical potential for the pointlike approximation in Fig.\ \ref{figpointlike}. The calculation
that leads to this phase diagram was first presented in Ref.\ \cite{Bergman:2007wp}, and we recapitulate it in appendix \ref{app:pointlike}. 
Besides the free energy of pointlike baryons, we also 
compute the free energy of the vacuum (= mesonic phase) and the chirally symmetric phase (= quark matter phase) in that appendix. The most important features of this phase diagram in our context are
the second-order transition from the vacuum to nuclear matter and the non-restoration of chiral symmetry for small temperatures and arbitrarily large chemical potential. 
In the rest of the paper we ask the question whether these two unphysical properties can be improved by going beyond the pointlike approximation.   

\subsection{All orders in non-abelian field strengths}
\label{sec:allorders}

We will now discuss a version of the DBI action in which we keep the non-abelian field strengths to all orders. For small chemical potentials (which correspond to small 
baryon densities) the results of this action are identical to the ones from the expanded action. For large chemical potentials, the results will obviously differ, and a priori it
is not clear which version -- if any -- is ``correct'', because there is no unique definition of the non-abelian DBI action to all orders in the string tension $\alpha'$.
We shall work with a non-expanded action mainly for two reasons. Firstly, the evaluation of the 
expanded Lagrangian (\ref{Lexp}) is challenging, and using the non-expanded Lagrangian discussed here is considerably simpler, see remarks below Eq.\ (\ref{da0dx4}). Secondly, we have checked that for the expanded 
Lagrangian there is -- within our ansatz -- no solution already beyond a moderately large chemical potential, while the solution for the non-expanded Lagrangian 
extends to very large chemical potentials, into a regime where the baryonic phase is already disfavored compared to the chirally restored phase. 

For the non-abelian DBI action we shall use the following prescription. Let us first suppose we 
work only with abelian field strengths $\hat{F}_{0U}$, $\hat{F}_{ij}$, $\hat{F}_{iU}$. 
Then, we obtain
\bea \label{DBIdeconf}
\sqrt{{\rm det}(g+2\pi\alpha'\hat{F})} &=& U^4\left(\frac{R}{U}\right)^{3/4}\left\{(2\pi\alpha')^2f_T\hat{F}_{iU}^2
+(1+u^3 f_T x_4'^2 -\hat{a}_0'^2)
\left[1+\left(\frac{R}{U}\right)^3\frac{(2\pi\alpha')^2\hat{F}_{ij}^2}{2}\right]\right. \non[2ex]
&&\hspace{2cm}\left.  + \left(\frac{R}{U}\right)^3\frac{(2\pi\alpha')^4f_T(\hat{F}_{ij}\hat{F}_{kU}\epsilon_{ijk})^2}{4}\right\}^{1/2} \, .
\eea
Following Ref.\ \cite{Rozali:2007rx}, we simply replace $\hat{F}_{iU} \to F_{iU}$ and $\hat{F}_{ij} \to F_{ij}$ and perform the trace of 
each term separately. Using the expressions in Eq.\ (\ref{Fus}), we arrive at the Lagrangian 
\be \label{Sdeconfu}
{\cal L} = u^{5/2}\sqrt{(1+g_1 +u^3f_Tx_4'^2-\hat{a}_0'^2)(1+ g_2)} - n_I\hat{a}_0(u)q(u)  \, ,
\ee
where we have abbreviated
\be
g_1(u) \equiv \frac{f_T(u)u^{1/2}}{3u_c^2\sqrt{f_c(u)}}n_Iq(u) \, , \qquad g_2(u) \equiv \frac{u_c^2\sqrt{f_c(u)}}{3u^{7/2}}n_Iq(u) \, .
\ee

Our goal is to solve the equations of motion for $x_4$ and $\hat{a}_0$ and determine the free energy of the resulting state.  We work in the grand-canonical ensemble, where the chemical potential is externally given and introduced
via the boundary value of $\hat{a}_0$,
\be \label{boundarya}
\hat{a}_0(\infty) = \mu \, .
\ee
To be precise, this is the quark chemical potential, such that $N_c\mu$ is the baryon chemical potential.
The boundary condition for $x_4$ is given by the asymptotic separation $L$ of the flavor branes,
\be \label{fixell}
\frac{\ell}{2} = \int_{u_c}^\infty du\, x_4' \, ,
\ee
with the dimensionless separation $\ell=M_{\rm KK} L$.

The equations of motion for $x_4$ and $\hat{a}_0$ in integrated form are  
\begin{subequations}
\bea
\frac{u^{5/2}\hat{a}_0'\sqrt{1+g_2}}{\sqrt{1+g_1+u^3f_Tx_4'^2-\hat{a}_0'^2}}&=& n_I Q \, , \label{eom1}\\[2ex]
\frac{u^{5/2}u^3f_Tx_4'\sqrt{1+g_2}}{\sqrt{1+g_1+u^3f_Tx_4'^2-\hat{a}_0'^2}}&=& k \, , \label{eom2}
\eea
\end{subequations}
where $k$ is an integration constant (to be determined), and we have denoted  
\be \label{Q}
Q(u) \equiv  \int_{u_c}^u dv\,q(v) = \frac{u^{3/2}\sqrt{f_c}}{2}\frac{3\rho^2u_c+2(u^3-u_c^3)}{(u^3-u_c^3+\rho^2u_c)^{3/2}}\, ,
\ee
such that $Q(\infty) = 1$, $Q(u_c)=0$. 
We can solve these equations algebraically for $\hat{a}_0'^2$ and $x_4'^2$,
\begin{subequations} \label{da0dx4}
\bea
\hat{a}_0'^2 &=&  \frac{(n_IQ)^2}{u^5}\frac{1+g_1}{1+g_2-\frac{k^2}{u^8f_T}+\frac{(n_IQ)^2}{u^5}} \, , \label{da0} \\[2ex]
x_4'^2 &=&  \frac{k^2}{u^{11}f_T^2}\frac{1+g_1}{1+g_2-\frac{k^2}{u^8f_T}+\frac{(n_IQ)^2}{u^5}} \, . \label{dx4}
\eea
\end{subequations}
At this point, the benefit of using the non-expanded Lagrangian (\ref{Sdeconfu}) becomes apparent: had we worked with the expansion in non-abelian gauge fields (\ref{Lexp}), 
$\hat{a}_0'^2$ and $x_4'^2$ would have been solutions to cubic equations, which is much more complicated to deal with (unless we had also employed an expansion in the abelian 
gauge fields). 

The asymptotic behavior at $u=\infty$ of the solutions is
\be \label{x4a0asymp}
x_4'(u) = \frac{k}{u^{11/2}} + \ldots \,, \qquad \hat{a}_0'(u) = \frac{n_I}{u^{5/2}} + \ldots\, , 
\ee
confirming that $n_I$ is the baryon density according to the usual AdS/CFT dictionary, which can also be checked later numerically by computing the derivative of the free energy with respect to $\mu$.  Using these solutions, we can write the free energy density $\Omega = \frac{T}{V} S$ in the useful forms
\bea \label{OmcalN}
\frac{\Omega}{\cal N} &=& \int_{u_c}^\infty du\left[(1+g_2)\frac{u^8f_Tx_4'}{k}+n_I\hat{a}_0'Q\right]-\mu n_I \non[2ex]
&=& \int_{u_c}^\infty du\,u^{5/2}\zeta\left[1+g_2+\frac{(n_IQ)^2}{u^5}\right] - \mu n_I \, ,
\eea
where we have introduced the abbreviation
\be \label{defzeta0}
\zeta \equiv \sqrt{\frac{1+g_1+u^3f_Tx_4'^2-\hat{a}_0'^2}{1+g_2}} = \frac{u^{11/2}f_Tx_4'}{k} = \frac{u^{5/2}\hat{a}_0'}{n_IQ} 
= \frac{\sqrt{1+g_1}}{\sqrt{1+g_2-\frac{k^2}{u^8f_T}+\frac{(n_IQ)^2}{u^5}}} \, . 
\ee
This free energy is divergent at the holographic boundary $u\to \infty$. Replacing the upper boundary of the integral by a cutoff $\Lambda$, one finds that the divergent
contribution is $2/7\,\Lambda^{7/2}$. All phases we consider in this paper have the same divergence, which is a pure vacuum contribution, i.e., does not depend
on $\mu$ or $T$. Therefore, and since we are only interested in differences of free energies, we can simply drop this contribution.

Besides the functions $x_4(u)$ and $\hat{a}_0(u)$, the system contains the parameters $n_I$, $\rho$, and $u_c$, and we have to minimize the free energy with respect to them.
At first sight it seems curious to minimize with respect to the density $n_I$, since we work in the grand-canonical ensemble where the density should be given as a function of $\mu$. 
However, in our setup $n_I$ should primarily be considered as a parameter of the Lagrangian. This parameter turns out to be identical
to the baryon density.
This is not necessarily the case, since the total density may receive contributions from other sources, for instance through a magnetic field \cite{Preis:2011sp}.

In order to minimize the free energy with respect to $n_I$, $\rho$, and $u_c$, it is convenient to read the Lagrangian as a functional of 
$\hat{a}_0(u),\hat{a}_0'(u),x_4'(u),n_I,\rho,u_c$, where the three functions  $\hat{a}_0(u),\hat{a}_0'(u),x_4'(u)$ also depend on $n_I,\rho,u_c$. 
Then, the minimization with respect to $n_I$ can be written as 
\be
0= \left(\frac{\partial{\cal L}}{\partial \hat{a}_0'}\frac{\partial \hat{a}_0}{\partial n_I}+\frac{\partial{\cal L}}{\partial x_4'}\frac{\partial x_4}{\partial n_I}
\right)_{u=u_c}^{u=\infty}+\int_{u_c}^\infty du\,\frac{\partial{\cal L}}{\partial n_I} \, , 
\ee
where we have used the equations of motion. All boundary terms vanish because the derivative is taken at fixed $\mu=\hat{a}_0(\infty)$ and fixed
separation $\ell/2=x_4(\infty) - x_4(u_c)$, and because
\be
\left.\frac{\partial{\cal L}}{\partial \hat{a}_0'}\right|_{u=u_c} = 0
\ee
due to Eq.\ (\ref{eom1}). Consequently, only the explicit derivative with respect to $n_I$ remains. The derivative with respect to the 
baryon width $\rho$ is taken completely analogously. The derivative with respect to $u_c$ is a bit more subtle. We obtain
\be \label{dOmduc1}
0=\left(\frac{\partial{\cal L}}{\partial \hat{a}_0'}\frac{\partial \hat{a}_0}{\partial u_c}+\frac{\partial{\cal L}}{\partial x_4'}\frac{\partial x_4}{\partial u_c}
\right)_{u=u_c}^{u=\infty}-{\cal L}\big|_{u=u_c}+\int_{u_c}^\infty du\,\frac{\partial{\cal L}}{\partial u_c} \, .
\ee 
Now we need to take into account that $\hat{a}_0(u_c)$ and $x_4(u_c)$ depend on $u_c$ explicitly and through the dependence on $u$. In the boundary terms, however,
only the explicit dependence is relevant. We thus write
\be
\left.\frac{\partial x_4}{\partial u_c}\right|_{u=u_c} =\frac{\partial x_4(u_c)}{\partial u_c} - x_4'(u_c) 
\ee
(and the same for $\hat{a}_0$), where the left-hand side is needed in Eq.\ (\ref{dOmduc1}), 
the first term on the right-hand side denotes the full dependence on $u_c$, and the second term the dependence via $u$. Using this 
relation and 
\be
\frac{\partial{\cal L}}{\partial x_4'} =k  \, , 
\ee
which follows from Eq.\ (\ref{eom2}), it turns out that the boundary term at $u=u_c$ gives a contribution $kx_4'(u_c)$, i.e., Eq.\ (\ref{dOmduc1}) becomes
\be
0 = (kx_4'-{\cal L})_{u=u_c} + \int_{u_c}^\infty du\,\frac{\partial {\cal L}}{\partial u_c} \, .
\ee
In summary, the three equations to minimize the free energy can be written as
\begin{subequations} \label{mini}
\bea
0 &=&  \int_{u_c}^\infty du\, \left[\frac{u^{5/2}}{2}\left(\frac{\partial g_1}{\partial n_I}\zeta^{-1}
+\frac{\partial g_2}{\partial n_I}\zeta \right) + \hat{a}_0'Q\right]-\mu \, , \label{mini1}\\[2ex]
0 &=&  \int_{u_c}^\infty du\, \left[\frac{u^{5/2}}{2}\left(\frac{\partial g_1}{\partial \rho}\zeta^{-1}
+\frac{\partial g_2}{\partial \rho}\zeta \right) + n_I \hat{a}_0'\frac{\partial Q}{\partial \rho}\right] \, ,\label{mini2} \\[2ex]
0 &=& -u_c^{5/2}\sqrt{\left[1+g_1(u_c)\right]\left[1+g_2(u_c)-\frac{k^2}{u_c^8f_T(u_c)}\right]}
+\int_{u_c}^\infty du\, \left[\frac{u^{5/2}}{2}\left(\frac{\partial g_1}{\partial u_c}\zeta^{-1}
+\frac{\partial g_2}{\partial u_c}\zeta \right) + n_I \hat{a}_0'\frac{\partial Q}{\partial u_c}\right] \, . \label{mini3}
\eea
\end{subequations}
In all three equations, we have eliminated $\hat{a}_0$ in favor of $\hat{a}_0'$ via 
partial integration. This is advantageous because $\hat{a}_0$ can only be obtained by numerically integrating Eq.\ (\ref{da0}), i.e., by eliminating $\hat{a}_0$ we avoid 
two nested numerical integrations.

We have arrived at a system of coupled algebraic equations -- Eqs.\ (\ref{mini}) plus the equation for the asymptotic separation (\ref{fixell}) --
which has to be solved for $k$, $n_I$, $u_c$, and $\rho$ for given chemical potential $\mu$ and temperature $T$ [which appears in $f_T(u)$]. 
Before discussing the results, let us comment on the numerical evaluation of these equations.

\subsection{Numerical evaluation}
\label{sec:numerical}

The equation that requires some explanation is the minimization with respect to $u_c$ (\ref{mini3}), for which we need to know the behavior of various functions at the tip of the connected flavor branes, $u=u_c$. We find
\be \label{expqQ}
n_Iq(u) = \frac{\sqrt{3}\alpha u_c^2}{\sqrt{u-u_c}} + {\cal O}[(u-u_c)^{1/2}] \,, \qquad n_IQ(u) = 2\sqrt{3}\alpha u_c^2\sqrt{u-u_c} + {\cal O}[(u-u_c)^{3/2}] \, , 
\ee
which implies 
\be
g_1(u) = \frac{\alpha f_T(u_c) u_c}{3(u-u_c)} + {\cal O}(1) \, , \qquad g_2(u)= \alpha + {\cal O}(u-u_c)  \, , 
\ee
where we have abbreviated
\be
\alpha\equiv \frac{3n_I}{4\rho u_c^{3/2}} \, .
\ee
Inserting these expansions into the solution for $x_4'$ (\ref{dx4}), we find that $x_4'$ diverges at $u=u_c$,
\be \label{x4uuc}
x_4'(u) = \frac{c_1}{\sqrt{u-u_c}} + {\cal O}[(u-u_c)^{1/2}] \, , \qquad c_1 \equiv \frac{1}{\sqrt{3}u_c}\frac{\alpha^{1/2} k}{\sqrt{u_c^8 f_T(u_c)(1+\alpha)-k^2}} \, .
\ee
Consequently, the embedding of the flavor branes $x_4(u)$ is smooth at $u=u_c$ for all values of $n_I$. This shows that the cusp, introduced by the approximation of the delta-like
baryons, is, not surprisingly, removed by the smooth instantonic baryons; see also appendix \ref{sec:pointlike}, where we review the pointlike approximation and see that 
in that case $x_4'$ is finite at $u=u_c$ and only diverges for $n_I\to 0$.

\begin{table*}[t]
\begin{tabular}{|c|c|c|c|c|c|c|c|} 
\hline
\rule[-1.5ex]{0em}{4ex} 
 $x_4'$ & $\;q,p,c_1\;$ & $x_4$ & $\;Q,\hat{a}_0',g_1,g_2,\alpha,\zeta\;$ & $\;u,u_T,\rho,\hat{a}_0,\mu\;$ & $n_I$ & $\;\Omega/{\cal N}\;$ & $k$ \\[1ex] \hline\hline
\rule[-1.5ex]{0em}{4ex} 
$\;\ell^3\;$ & $\ell^2$ & $\;\ell\;$ & $1$ & $\ell^{-2}$ & $\;\ell^{-5}\;$ & $\;\ell^{-7}\;$ &  $\;\ell^{-8}\;$ 
\\[1ex] \hline
\rule[-1.5ex]{0em}{4ex} 
$\;u_c^{-3/2}\;$ & $u_c^{-1}$ & $\;u_c^{-1/2}\;$ & $1$ & $u_c$ & $\;u_c^{5/2}\;$ & $\;u_c^{7/2}\;$ &  $\;u_c^4\;$ 
\\[1ex] \hline
\end{tabular}
\caption{The system of four coupled equations (\ref{fixell}), (\ref{mini1}) -- (\ref{mini3}) is most conveniently solved by first rescaling all quantities with the 
constant $\ell$, and then with the variable $u_c$, as given in this
table. As a consequence, $\ell$ disappears from the equations, and $u_c$ only remains in Eq.\ (\ref{fixell}), which thus decouples from the other three equations. 
}
\label{table1}
\end{table*} 

We can now compute 
\be \label{Luc}
-u_c^{5/2}\sqrt{\left[1+g_1(u_c)\right]\left[1+g_2(u_c)-\frac{k^2}{u_c^8f_T(u_c)}\right]} = -\frac{\alpha k}{3u_c^2c_1 \sqrt{u-u_c}}+ {\cal O}[(u-u_c)^{1/2}] \, ,
\ee
i.e., we have obtained a divergent contribution (and no constant term). However, the integral in Eq.\ (\ref{mini3}) also contains a divergent term, and both divergences 
exactly cancel each other. The divergent term of the integral arises in 
\be
\frac{\partial g_1}{\partial u_c}\zeta^{-1}+\frac{\partial g_2}{\partial u_c}\zeta = g_1(u)\left[p(u)-\frac{2}{u_c}\right]\zeta^{-1}(u)+g_2(u)\left[p(u)+\frac{2}{u_c}\right]\zeta(u)
-\frac{g_1(u)}{f_c(u)}\frac{\partial f_c}{\partial u_c}\zeta^{-1}(u) \, , 
\ee
where we have abbreviated the function
\be \label{pu}
p(u) \equiv \frac{2}{u_c}+\frac{5}{2}\frac{3u_c^2-\rho^2}{u^3-u_c^3+\rho^2u_c} \, ,
\ee
which originates from taking the derivative of $q(u)$ with respect to $u_c$. Now, with 
\be
\frac{\partial Q}{\partial u_c} = -q(u)\frac{u^3+2 u_c^3}{3u^2u_c} \, , 
\ee
we can write Eq.\ (\ref{mini3}) as
\bea \label{minuc}
0 &=& \int_{u_c}^\infty du 
\left[u^{5/2}\frac{\zeta g_2\left(p+\frac{2}{u_c}\right)+\zeta^{-1}g_1\left(p-\frac{2}{u_c}\right)}{2}-n_I q \hat{a}_0'\frac{u^3+2u_c^3}{3u^2u_c}-\frac{\alpha k}{6u_c^2 c_1(u-u_c)^{3/2}}
+\frac{3u_c^2}{u^{1/2}f_c}\frac{\zeta^{-1}g_1}{2} 
\right] \, , 
\eea
and one easily checks that the divergences in the two last terms cancel each other, rendering the integral finite. The other equations (\ref{fixell}), (\ref{mini1}), and (\ref{mini2}),
are evaluated much more straightforwardly and require no further explanation.

\begin{figure} [t]
\begin{center}
\underline{Confined geometry, instanton gas}

\vspace{0.2cm}
\hbox{\includegraphics[width=0.5\textwidth]{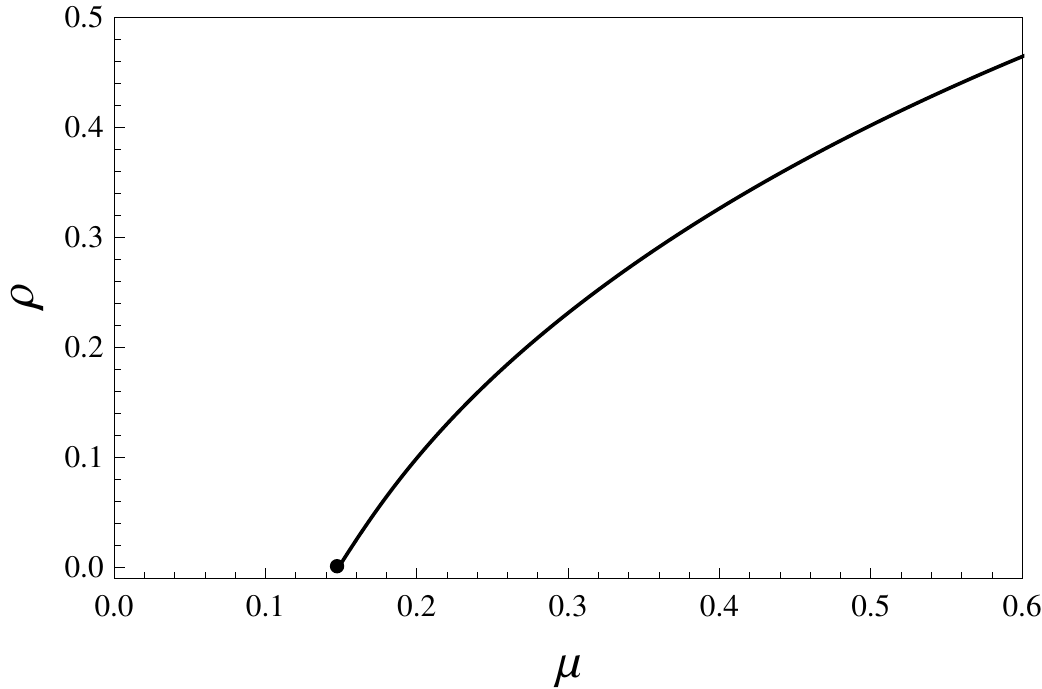}\includegraphics[width=0.5\textwidth]{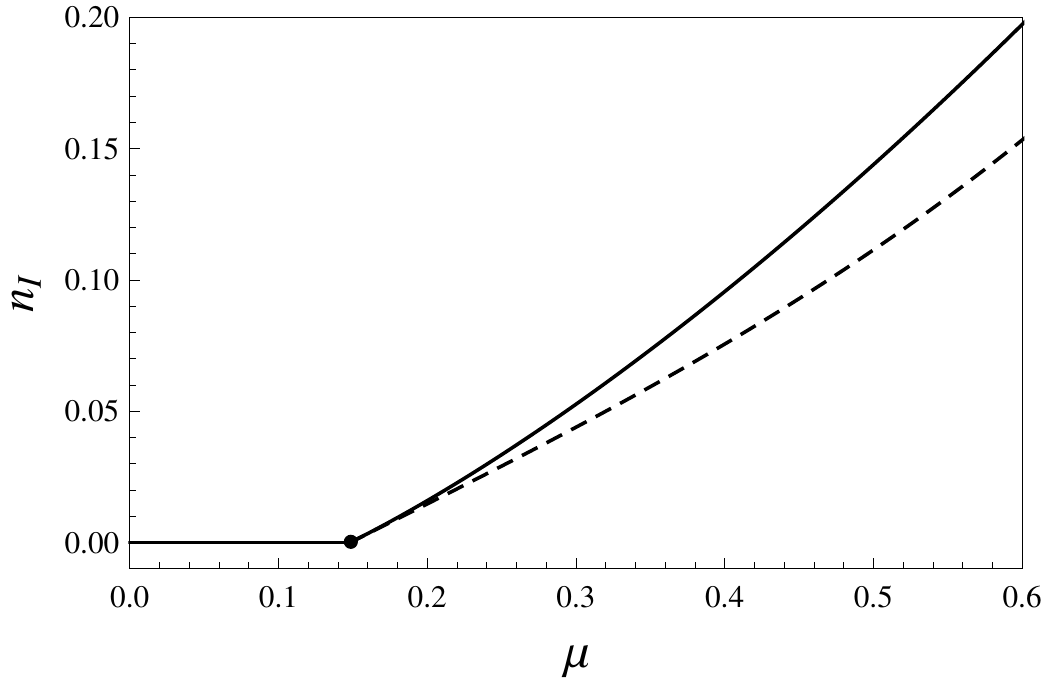}}
\hbox{\includegraphics[width=0.5\textwidth]{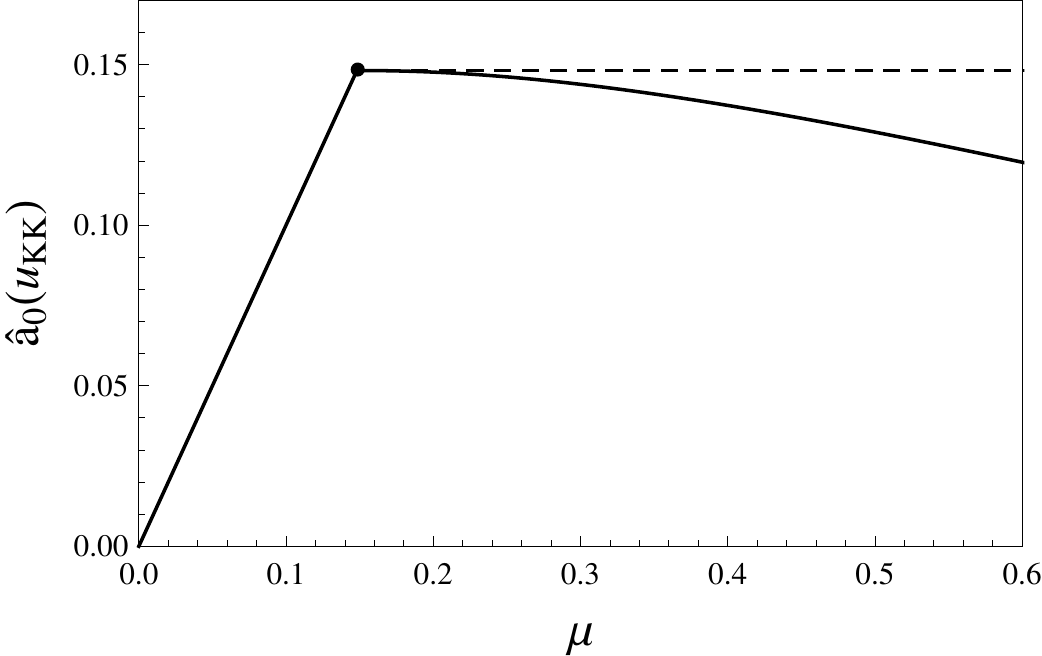}\includegraphics[width=0.5\textwidth]{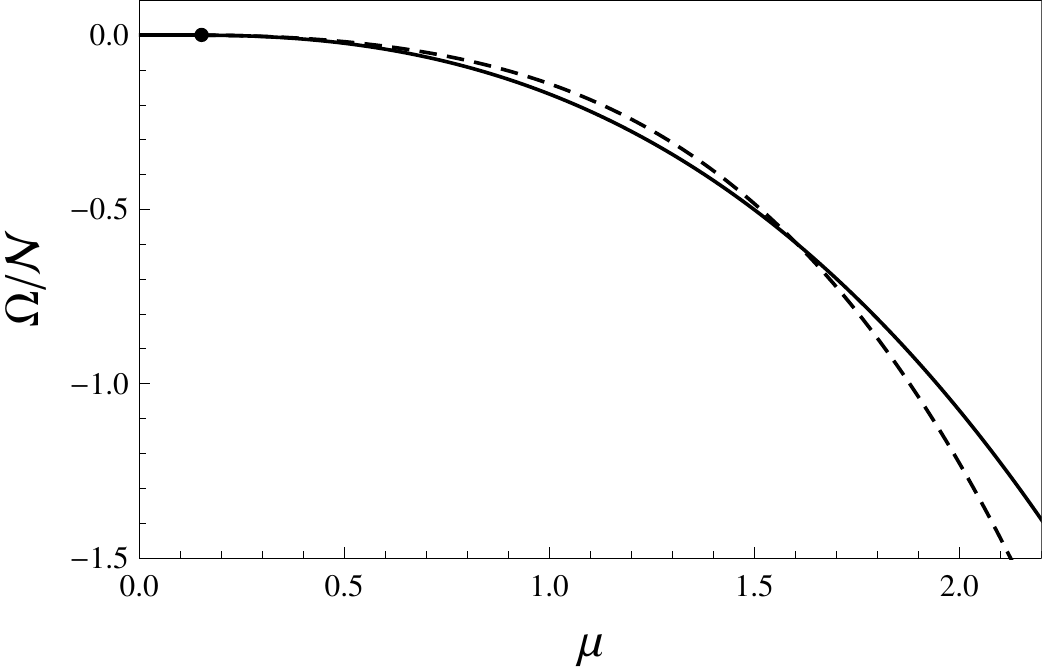}}
\caption{Confined geometry 
with maximally separated flavor branes: baryon width $\rho$ (upper left panel), baryon density $n_I$ (upper right panel),
value of $\hat{a}_0$ at the tip of the connected flavor branes (lower left panel), and free energy density (lower right panel), 
all as a function of the chemical potential $\mu$. (Note the larger $\mu$ scale in the lower right panel.)
The dashed lines correspond to the results for pointlike baryons, where $\rho=0$. The plots show the second-order baryon onset 
at $\mu = u_{\rm KK}/3 = 4/27 \simeq 0.148$. Below the onset, i.e., in the mesonic phase, $\hat{a}_0(u_{\rm KK})=\mu$, $n_I=0$, $\Omega=0$, and $\rho$ undefined.
For the dimensionless units used here and in all following plots, see Table \ref{table0}.}
\label{figconf1}
\end{center}
\end{figure} 

For the numerical evaluation, we first note that rescaling of all quantities with appropriate powers of $\ell$ eliminates $\ell$ from all equations. The various powers
of $\ell$ are given in Table \ref{table1}. We shall not introduce new symbols for the quantities rescaled with $\ell$, but rather write $\ell$ explicitly in all plots.  
Then, we can proceed analogously with $u_c$ (remember that $\ell$ is a constant parameter, while $u_c$ is a dynamically determined
variable). Firstly, this removes the 
variable $u_c$ from the lower boundary of the integrals, which is convenient for the numerical evaluation. 
And secondly, after this rescaling, $u_c$ drops completely out of Eqs.\ (\ref{mini1}) -- (\ref{mini3}), i.e., 
we can solve them for (the rescaled) $k$, $n_I$, $\rho$ and afterwards compute $u_c$ from Eq.\ (\ref{fixell}), which is then used to undo the rescaling of 
$k$, $n_I$, $\rho$. 

We also note that $\mu$ only appears in Eq.\ (\ref{mini1}), and thus 
one can further reduce the number of coupled equations to two if one fixes $n_I$ and solves for $k$ and $\rho$, and afterwards computes the resulting $\mu$ and $u_c$. Finally, we mention
that the numerical integration is best done after a change of variables according to Eq.\ (\ref{Z}) from $u$ to $z$ with $z$ defined as $u=(1+z^2)^{1/3}$ (after scaling out $u_c$). 
This is particularly helpful for the numerical integration in Eq.\ (\ref{minuc}), which is the most challenging one.

\subsection{Second-order baryon onset and chiral restoration}
\label{sec:2nd}

In this section, we present our numerical results for the instanton gas. Our emphasis is on the deconfined geometry. Nevertheless, we start with the confined geometry
with maximally separated flavor branes. This allows us to compare our results to the literature and serves as a warm-up exercise for the more difficult 
calculations in the deconfined geometry. 
All relevant equations for the confined geometry are collected in appendix \ref{app:confinst}; the calculation 
basically reduces to solving Eqs.\ (\ref{miniconf}), the stationarity equations for the baryon density $n_I$ and the baryon width $\rho$. In contrast to the deconfined geometry, 
the results do not depend on temperature. The location of the tip of the connected flavor branes is fixed in the case of 
maximally separated branes and identical to $u_{\rm KK}$. Baryon width, baryon density, $\hat{a}_0(u_{\rm KK})$, and the free energy density are plotted as a function of 
the chemical potential in Fig.\ \ref{figconf1}. We see that for small $n_I$ we reproduce the 
results of the pointlike approximation, in particular we find a second-order baryon onset\footnote{\label{footGhoroku}In Ref.\ \cite{Ghoroku:2012am} it was claimed in an almost identical 
calculation that the baryon onset becomes first order within the approximation of the instanton gas. Even though in that reference the DBI action was expanded for small 
non-abelian gauge fields, the transition to baryonic matter should be of the same order because our result reduces to that of the expanded action for small baryon densities.
However, in Ref.\ \cite{Ghoroku:2012am}, $\hat{a}_0(u_{\rm KK})$ 
has been set to zero. In doing so, a lower boundary for the baryon density is ``forced'' upon the system (and the free energy is minimized only with respect to $\rho$, 
not also with respect to $n_I$). Our calculation shows that if $\hat{a}_0(u_{\rm KK})$ is determined dynamically, it chooses to approach $\mu$ at the baryon 
onset, connecting continuously to the mesonic phase, where $\hat{a}_0(u)=\mu$ for all $u$, see lower left panel of Fig.\ \ref{figconf1}. 
As a consequence, the baryon density vanishes at that point and the 
onset is second order.}.  The onset occurs at $\mu = u_{\rm KK}/3 = 4/27$. With the help of Table \ref{table0} and the 
comments below Eq.\ (\ref{ED4}) we see that this critical chemical potential is identical to the baryon mass, which implies that there is no binding energy. 
The plot of the free energy shows that at very large chemical potentials the finite-size baryons become 
energetically more costly than pointlike baryons. At even higher chemical potentials, $\mu\gtrsim 2.37$, beyond the scale shown in Fig.\ \ref{figconf1},
we do not find any solution. (There is a second solution up to that chemical potential $\mu \simeq 2.37$ which we have not included in the plots since its free energy 
is always higher than the one of the solution plotted here).

\begin{figure} [t]
\begin{center}
\underline{Deconfined geometry, instanton gas, $T=0$}

\vspace{0.2cm}
\hbox{\includegraphics[width=0.5\textwidth]{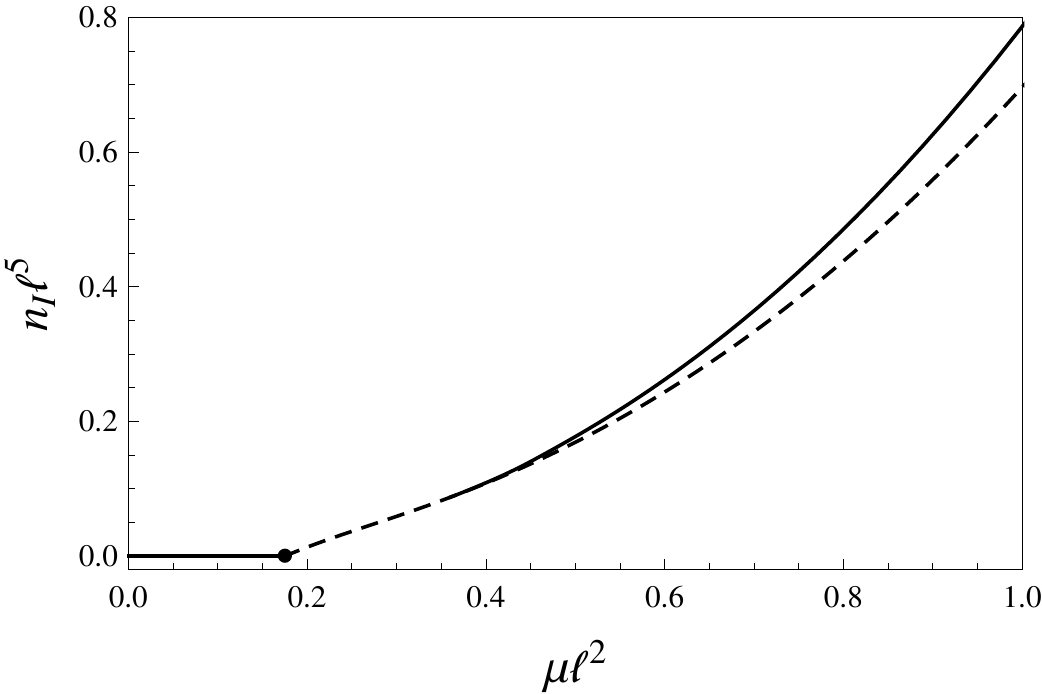}\includegraphics[width=0.5\textwidth]{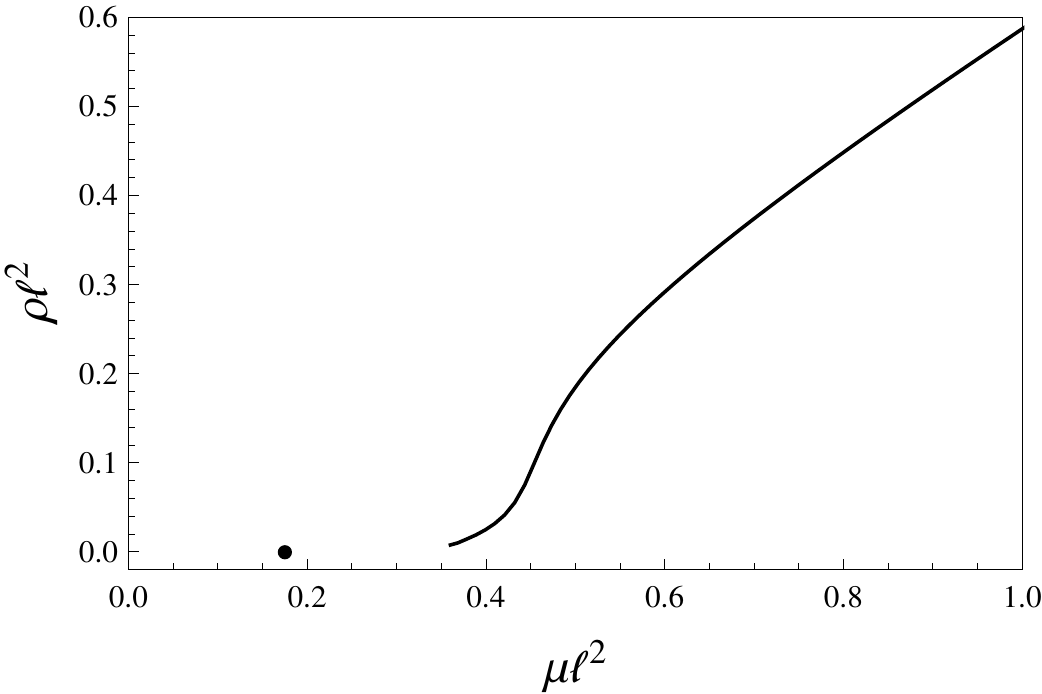}}
\hbox{\includegraphics[width=0.495\textwidth]{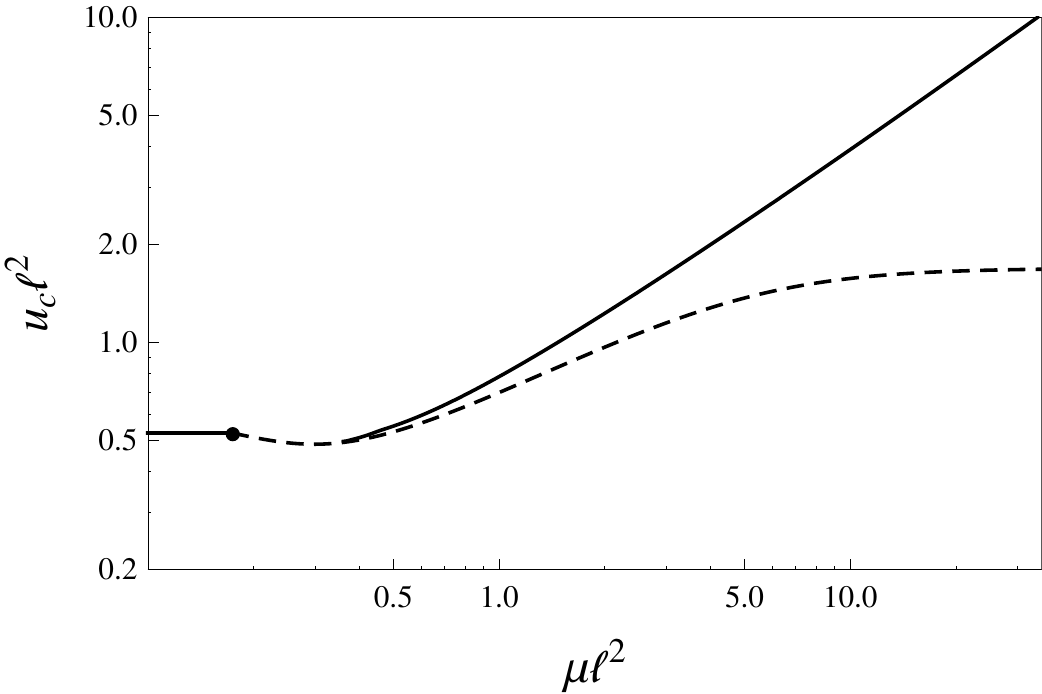}\hspace{0.2cm}\includegraphics[width=0.495\textwidth]{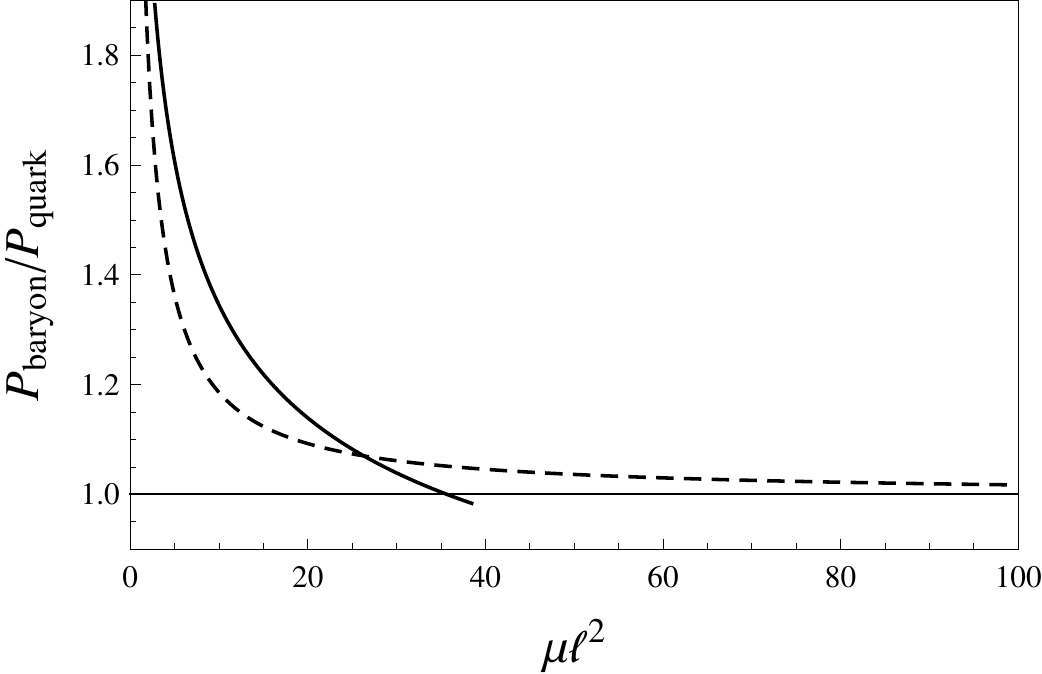}}
\caption{Deconfined geometry at zero temperature: baryon density $n_I$ (upper left panel), 
baryon width $\rho$ (upper right panel), location of the tip of the connected flavor branes $u_c$ (lower left panel), and ratio of the pressures of baryonic and quark matter
phases (lower right panel), all as a function of $\mu$ and rescaled with the asymptotic separation of the flavor branes $\ell$. 
(Note the larger $\mu$ scale in the two lower panels -- logarithmic in the lower left panel.)
The dashed lines are the results for pointlike baryons. Due to numerical difficulties for very small $\rho$, the solid lines do not connect with 
the lines of the mesonic phase below the second-order baryon onset. 
In the lower right panel, the solid line stops at a point beyond which we have not found any solutions (and where baryonic matter is already disfavored).
}
\label{figdeconf1}
\end{center}
\end{figure}

\begin{figure} [t]
\begin{center}
\includegraphics[width=0.5\textwidth]{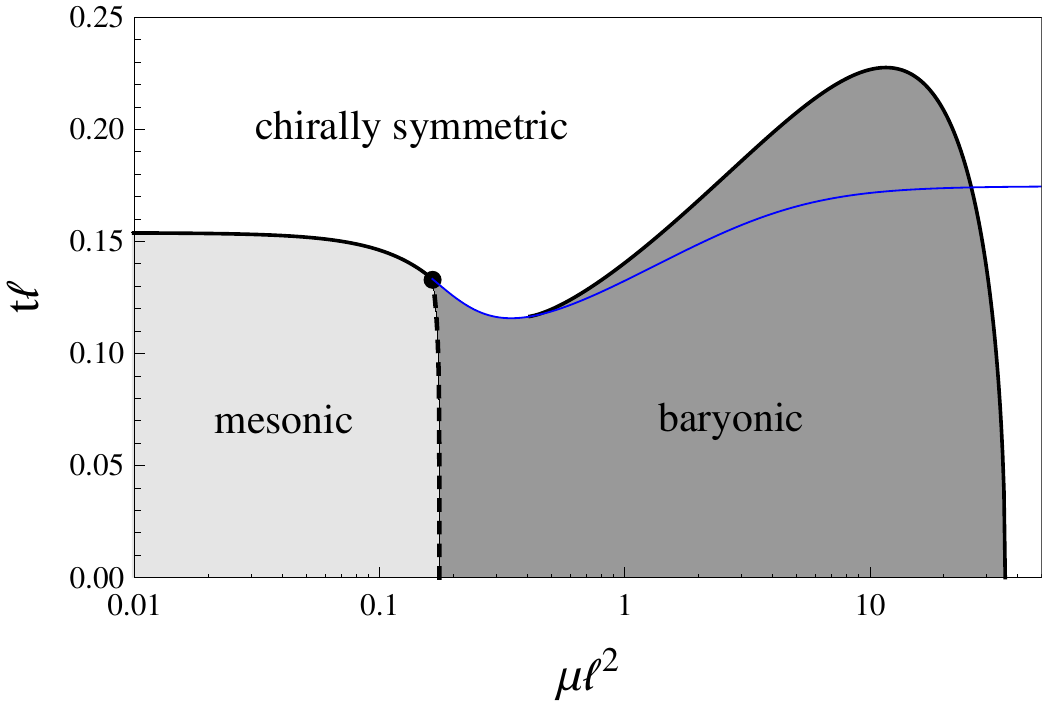}
\caption{Phase diagram in the deconfined geometry with baryons in the instanton gas approximation. Since for small baryon density the result approaches the one from pointlike baryons,
the second-order baryon onset (as well as the transition from the mesonic to the chirally restored phase) are identical to the ones in Fig.\ \ref{figpointlike}.
The first-order transition between baryonic and chirally symmetric phases is shown by the thin (blue) line for pointlike baryons and by the thick line for instantonic baryons. 
The main qualitative difference is the chiral restoration for small temperatures at very large $\mu$ (note the logarithmic $\mu$ scale, while the $\mu$ scale in Fig.\ \ref{figpointlike}
is linear). The black line for the instanton gas 
is not connected to the critical point where the second-order onset line meets the first-order chiral transition line because of numerical difficulties 
at small baryon width, see also Fig.\ \ref{figdeconf1}.} 
\label{figphasesinst}
\end{center}
\end{figure}

We now turn to the deconfined geometry, whose results are shown in Figs.\ \ref{figdeconf1} (zero temperature) and \ref{figphasesinst} (phase diagram). 
In the upper left panel of Fig.\ \ref{figdeconf1} we show the 
baryon density $n_I$ and compare it to that of the pointlike approximation. As for the confined geometry, we observe a second-order phase transition whose properties around the 
transition (i.e., for small $n_I$) are well approximated by pointlike baryons. This transition is reminiscent of a Bose-Einstein condensation \cite{Preis:2011sp}, which 
is maybe not surprising since at infinitely large $N_c$ baryons may lose their $N_c=3$ fermionic nature. 
Our numerical result for the instanton gas is not shown all the way
down to the onset because the numerical evaluation becomes problematic at very small baryon densities. This regime corresponds to very small baryon widths $\rho$, as the upper right panel
shows, and thus we have to integrate numerically over highly peaked functions, which is hard. However, the plots show that our calculation works well down to 
a regime where
the result already merges with the pointlike approximation. And, in the confined geometry, where we {\it can} perform the calculation down to the onset, we have seen that 
instanton gas and pointlike approximation become indeed identical. Therefore, it would be very surprising if the full result in the deconfined case did not follow this approximation down to the baryon onset, even though our numerics cannot prove it.

The lower left panel shows the location of the tip of the connected flavor branes $u_c$. This is an interesting quantity because it can be related to the medium-dependent 
baryon mass, at least in the pointlike approximation, see Eq.\ (\ref{ED4}) and comments below that equation. Interestingly, the behavior of $u_c$ for the instanton gas at large chemical potentials differs
qualitatively from that of the pointlike approximation. While instantonic baryons appear to become heavier without bound, the pointlike baryon mass saturates at a finite value. As a consequence, one can expect the instanton gas to become energetically more costly with increasing $\mu$ relative to the gas of pointlike baryons. 
This expectation is borne out in the 
lower right panel: the pressure $P=-\Omega$ of pointlike baryons approaches that of the chirally restored phase, $P\propto \mu^{7/2}$, 
from above \cite{Preis:2011sp}, thus being favored over quark matter for arbitrarily large $\mu$. (The free energy of the chirally restored phase 
is derived in appendix \ref{sec:restored}.)
In contrast, while the instanton gas is energetically {\it less} costly (= has higher pressure) 
at small $\mu$ than the pointlike baryons, the curve of its pressure crosses the pressure curves of the 
pointlike approximation {\it and} the one of chirally restored quark matter. 
This shows that in the instanton gas approximation there is a zero-temperature chiral phase transition to quark matter at large chemical potential.
The corresponding critical chemical potential is extremely large, about 200 times larger than the chemical potential for the baryon onset 
($\mu \simeq 35/\ell^2$, while the onset 
occurs at $\mu\simeq 0.17/\ell^2$). Therefore, if we were to fit the baryon onset to QCD (which makes little sense due to its obviously unrealistic properties), our calculation 
would predict chiral restoration to occur at a baryon chemical potential of about $200\,{\rm GeV}$. We have to keep in 
mind, however, that the result at large $\mu$ has to be taken with a lot of care because of various reasons:  we have used a simple, not uniquely defined, prescription of the non-abelian DBI action, whose validity is guaranteed only for sufficiently small $\mu$; we have employed the probe brane approximation, which becomes questionable for very large gauge fields on the flavor branes;  and, the instantons start to overlap at densities well below the critical chemical potential for the chiral phase transition, which challenges our instanton
gas approximation.  
For these reasons we do not attempt to use this approximation for any quantitative predictions, since clearly it first has to be improved in the future.
Nevertheless, the existence of chiral restoration of baryonic matter in the Sakai-Sugimoto model at high densities, at least within a certain approximation, 
is a very interesting and -- to our knowledge -- new observation and important progress towards a high-density, strongly coupled model 
outlined in the introduction.

We can now compute the phase transition between baryonic and chirally restored phases for all temperatures. The resulting phase diagram is shown in 
Fig.\ \ref{figphasesinst}.

\section{Baryonic matter from a homogeneous ansatz}
\label{sec:hom}

\subsection{Ansatz and calculation}
\label{sec:ancalc}

So far, we have worked with an ansatz for the non-abelian gauge fields that was based on the BPST instanton solution. In this section, we discuss a second ansatz,
given by \cite{Rozali:2007rx}
\be \label{ansatzhom}
A_Z = 0 \, , \qquad A_i(Z) = -\sigma_i\frac{H(Z)}{2}  \, ,
\ee 
with some function $H(Z)$, to be determined. With $F_{\mu\nu} = \partial_\mu A_\nu - \partial_\nu A_\mu + i[A_\mu,A_\nu]$, the field strengths become
\be
F_{ij} = -\epsilon_{ijk}\sigma_k\frac{H^2}{2} \, , \qquad F_{iZ} = \sigma_i\frac{\partial_Z H}{2} \, ,
\ee 
where $[\sigma_i,\sigma_j]=2i\epsilon_{ijk}\sigma_k$ has been used.
This ansatz is different from the instanton solution (\ref{AzAiinst}), where both $A_Z$ and $A_i$ are nonvanishing and depend not only on the holographic coordinate
$Z$ but also on position space. In our approximation of the previous section, we have integrated out the position space dependence before solving the equations of 
motion, i.e., eventually also our instanton gas was spatially homogeneous. Nevertheless, we shall refer to Eq.\ (\ref{ansatzhom}) as ``homogeneous'' since it does not 
involve any spatial dependence to begin with. As noticed already in 
Ref.\ \cite{Rozali:2007rx}, the ansatz (\ref{ansatzhom}) only yields a nonzero baryon density if $A_i(Z)$ is allowed to become discontinuous at $Z=0$. 
The reason is that the baryon density assumes the form 
\be \label{NH3}
-\frac{1}{8\pi^2}\int_{-\infty}^\infty dZ\,\Tr[F_{ij}F_{kZ}]\epsilon_{ijk} = \frac{1}{8\pi^2}\int_{-\infty}^\infty dZ\,\partial_Z H^3  \, , 
\ee
and $H(\pm \infty)=0$ has to be required in order to ensure a finite free energy (up to a constant divergent term that we always subtract). Therefore, if $H^3$ was a continuous
function, the integral (\ref{NH3}) would simply give $H(+\infty)^3-H(-\infty)^3=0$. 
We shall briefly return to this issue below Eq.\ (\ref{hexp}), when we know the exact behavior of $H(Z)$ around $Z=0$.

As for the instanton gas, we shall work in the coordinate $U$ and on one half of the connected flavor branes [for the relation between $Z$ and $U$ see Eq.\ (\ref{Z})]. 
With $\Tr[\sigma_i\sigma_i] = 6$ we have
\be
\Tr[F_{iU}^2] = \frac{3h'^2}{2(2\pi\alpha')^2}  \, , \qquad   \frac{\Tr[F_{ij}^2]}{(M_{\rm KK} R)^6} = \frac{3\lambda_0^2h^4}{(2\pi\alpha')^2} \, , \qquad 
\frac{\Tr[F_{ij}F_{kU}]\epsilon_{ijk}}{(M_{\rm KK} R)^3} = -\frac{3\lambda_0 h^2h'}{(2\pi\alpha')^2} \, , 
\ee
where the dimensionless gauge field $h$ is defined analogously to $\hat{a}_0$, see Table \ref{table0}, and where
we have abbreviated
\be
\lambda_0\equiv \frac{\lambda}{4\pi} \, .
\ee
Replacing all abelian $\hat{F}^2$ terms in Eq.\ (\ref{DBIdeconf}) with these
non-abelian versions and using the CS term (\ref{SCS}) yields the action 
\be \label{Shom}
S = {\cal N} \int_{u_c}^\infty du\,\left[u^{5/2}\sqrt{\left(1+\frac{3f_Th'^2}{2}+u^3f_T x_4'^2-\hat{a}_0'^2\right)\left(1+\frac{3\lambda_0^2h^4}{2u^3}\right)}
-\hat{a}_0\lambda_0\frac{9h^2h'}{2}\right] \, .
\ee
In contrast to the instanton gas, the Lagrangian now depends on the 't Hooft coupling explicitly, there is no rescaling of the fields by which we can get rid of $\lambda$. The
reason is the asymmetric ansatz for $A_Z$ and $A_i$, which leads to a different scaling behaviour of $F_{ij}$ and $F_{iU}$. Even though this ansatz is somewhat more simplistic 
than the instanton gas, the explicit appearance of $\lambda$ allows us to capture some physics away from the $\lambda=\infty$ limit. We shall see that only
at finite $\lambda$ we obtain a realistic first-order onset of baryonic matter. Of course, since we do not go beyond the classical gravity approximation, we cannot claim 
to include finite-$\lambda$ corrections systematically and have to consider our results at finite $\lambda$ as an extrapolation. 

The Lagrangian now contains the additional unknown function $h(u)$, besides $\hat{a}_0(u)$ and $x_4(u)$. We thus have to solve an additional Euler-Lagrange equation.
This seems to render the calculation more tedious, but we shall see that the three differential equations can be decoupled, and eventually the effort needed to evaluate this
approach numerically is comparable to the one needed for the instanton gas.   
Employing the same compact notation as in Sec.\ \ref{sec:allorders}, we abbreviate
\be
g_1(u) \equiv \frac{3f_Th'(u)^2}{2} \, , \qquad g_2(u) \equiv \frac{3\lambda_0^2h(u)^4}{2u^3} \, .
\ee
The baryon density is obtained from the CS term,
\be \label{nIh}
n_I = \frac{3\lambda_0}{2}\int_{u_c}^\infty du\, \partial_u h^3 = -\frac{3\lambda_0}{2}h(u_c)^3 \, , 
\ee
where we have used $h(\infty)=0$. (For the sake of a consistent notation, we continue to denote the baryon density by $n_I$, although the current approach is
not based on the instanton solution.)

The equations of motion for $\hat{a}_0$ and $x_4$ are 
\begin{subequations}
\bea
\frac{u^{5/2}\hat{a}_0'\sqrt{1+g_2}}{\sqrt{1+g_1+u^3f_T x_4'^2-\hat{a}_0'}} &=& n_I Q \, , \\[2ex] 
\frac{u^{5/2}u^3 f_T x_4'\sqrt{1+g_2}}{\sqrt{1+g_1+u^3f_T x_4'^2-\hat{a}_0'}} &=& k \, ,  
\eea
\end{subequations}
with the integration constant $k$ and, in analogy to the function $Q$ from Eq.\ (\ref{Q}), 
\be
Q(u)\equiv \frac{3\lambda_0}{2n_I}[h(u)^3-h(u_c)^3] = 1+\frac{3\lambda_0}{2n_I}h(u)^3 \, .
\ee
(To emphasize the analogy to the calculation of Sec.\ \ref{sec:inst} we use the same symbols $g_1$, $g_2$, $Q$, but of course need to keep in mind that they denote 
different functions.) In complete analogy to Eqs.\ (\ref{da0dx4}) we thus find
\begin{subequations}
\bea
\hat{a}_0'^2 &=& \frac{(n_IQ)^2}{u^5}\frac{1+g_1}{1+g_2-\frac{k^2}{u^8f_T}+\frac{(n_IQ)^2}{u^5}} \, , \\[2ex]
x_4'^2 &=& \frac{k^2}{u^{11}f_T^2}\frac{1+g_1}{1+g_2-\frac{k^2}{u^8f_T}+\frac{(n_IQ)^2}{u^5}} \, .\label{x4homdeconf}
\eea
\end{subequations}
For the equation of motion for $h$ we need
\be \label{dLdh}
\frac{\partial {\cal L}}{\partial h} = \frac{u^{5/2}\zeta}{2}\frac{\partial g_2}{\partial h} -9\lambda_0\hat{a}_0h'h \, , \qquad 
\frac{\partial {\cal L}}{\partial h'} = \frac{u^{5/2}\zeta^{-1}}{2}\frac{\partial g_1}{\partial h'}-\frac{9\lambda_0}{2}\hat{a}_0h^2 \, ,
\ee
where
\be
\zeta \equiv \frac{\sqrt{1+g_1+u^3f_T x_4'^2-\hat{a}_0'^2}}{\sqrt{1+g_2}} = \frac{\sqrt{1+g_1}}{\sqrt{1+g_2-\frac{k^2}{u^8f_T}+\frac{(n_IQ)^2}{u^5}}} \, .
\ee
The expression on the right-hand side does not contain $\hat{a}_0$ and $x_4$ anymore, which can thus be eliminated from the equation of motion for $h$,
\be  \label{eqh1}
\partial_u\left(u^{5/2}\zeta^{-1}\frac{\partial g_1}{\partial h'}\right)-\frac{9\lambda_0h^2\zeta n_IQ}{u^{5/2}} = u^{5/2}\zeta\frac{\partial g_2}{\partial h} \, .
\ee
We are thus left with this single differential equation for $h$ which has to be solved numerically. 
From Eq.\ (\ref{nIh}) we know the boundary value at $u=u_c$ in terms of $n_I$. By expanding Eq.\ (\ref{eqh1}) around $u=u_c$, we find the behavior
\bea \label{hexp}
h(u) &=& -\left(\frac{2n_I}{3\lambda_0}\right)^{1/3} + a_1\sqrt{u-u_c} + a_2 (u-u_c) + \ldots  \, , 
\eea
with coefficients $a_1$, $a_2$. Inserting this expansion into Eq.\ (\ref{x4homdeconf}), we see that $x_4'\propto 1/\sqrt{u-u_c}$ for $u$ close to $u_c$. This divergent 
derivative shows that the embedding of the flavor branes is smooth, just like for the instanton gas, i.e., there is no cusp at the tip of the connected flavor branes,
which occurs in the pointlike approximation. Remember from the discussion at the beginning of this section that the function $h(z)$ must be discontinuous at the tip. Thus, 
if we want to continue $h(z)$ into the second half of the connected flavor branes, we have to add a factor ${\rm sgn}(z)$, such that $h$ is antisymmetric under $z\to -z$. 
As a consequence, $\partial_z h^3$ is symmetric and continuous, and the $z<0$ half of the connected branes gives the same contribution 
to the baryon density as the $z>0$ half. The function $\partial_z h^3$ is the analogue of the instanton profile $q(z)$, see Eq.\ (\ref{qZ}). In contrast to the
instanton profile, $\partial_z h^3$ has a cusp at $z=0$. But unlike the delta-like baryons from the pointlike approximation, $\partial_z h^3$
is finite at $z=0$ and has a finite width. Having this picture of the complete connected flavor branes in mind, we can now go back to working on one single half.

For a given $a_1$, we can compute $k$: from the requirement that Eq.\ (\ref{eqh1}) be fulfilled to lowest order, $(u-u_c)^0$, we find 
\be
k^2= u_c^8f_T(u_c)\frac{8u_c-3a_1^2\left(5-2\frac{u_T^3}{u_c^3}\right)+\frac{3\lambda_0^2h(u_c)^4}{2u_c^3}
\left[8u_c-3a_1^2\left(2+\frac{u_T^3}{u_c^3}\right)\right]}{8u_c+9a_1^2f_T(u_c)} \, .
\ee
(The higher-order terms contain the higher-order coefficients $a_2$ etc., which can be computed successively as a function of $a_1$ in this way, but will not be needed in the following.)
We can thus eliminate $k$ in favor of $a_1$. For a given baryon density $n_I$, which determines the boundary value $h(u_c)$, 
we determine the coefficient $a_1$ numerically via the shooting method, such that $h(\infty)=0$. 

Next, we need to take into account the minimization of the free energy with respect to $n_I$. Reading the Lagrangian as a functional of $\hat{a}_0$, $\hat{a}_0'$, $x_4'$, $h$, $h'$
(with no further explicit dependence on $n_I$), we obtain 
\bea \label{dOmdnIhom}
0  &=& \left.\frac{\partial {\cal L}}{\partial \hat{a}_0'}\frac{\partial \hat{a}_0}{\partial n_I}\right|_{u=u_c}^{u=\infty} 
+\left.\frac{\partial {\cal L}}{\partial x_4'}\frac{\partial x_4}{\partial n_I}\right|_{u=u_c}^{u=\infty}
+\left.\frac{\partial {\cal L}}{\partial h'}\frac{\partial h}{\partial n_I}\right|_{u=u_c}^{u=\infty} \quad \Rightarrow \qquad 
\left.\frac{\partial {\cal L}}{\partial h'}\right|_{u=u_c} = 0 \, ,
\eea
where we have used that $\hat{a}_0(\infty)=\mu$, $x_4(\infty) - x_4(u_c)=\ell/2$, $h(\infty)=0$ are fixed and thus do not depend on $n_I$, and 
$\frac{\partial {\cal L}}{\partial \hat{a}_0'}|_{u=u_c} = 0$, $\frac{\partial {\cal L}}{\partial x_4'}|_{u=u_c} = k$.

With the help of Eq.\ (\ref{dLdh}), the condition (\ref{dOmdnIhom}) can be used to compute $\hat{a}_0(u_c)$,
\be \label{a0uc2}
\hat{a}_0(u_c) = \frac{u_c\sqrt{f_T(u_c)}}{3}\sqrt{1+\frac{2u_c^3}{3\lambda_0^2h^4(u_c)}\left[1-\frac{k^2}{u_c^8 f_T(u_c)}\right]}\, .
\ee
From this relation we can already see that for large 't Hooft couplings, $\lambda_0\to \infty$, we recover the result of pointlike baryons, see Eq.\ (\ref{a0uc}). 
We shall come back to this observation in our discussion of the numerical results. 

Recall that in the case of the instanton gas we additionally had to minimize with respect to the instanton width $\rho$ and the location of the tip of the connected flavor branes $u_c$. 
Now, there is no parameter $\rho$, and the minimization with respect to $u_c$ is automatically included in the equations of motion, which can be seen as follows.
Minimizing the free energy with respect to $u_c$ yields
\bea
0 &=& \left.\frac{\partial {\cal L}}{\partial \hat{a}_0'}\frac{\partial \hat{a}_0}{\partial u_c}\right|_{u=u_c}^{u=\infty} 
+\left.\frac{\partial {\cal L}}{\partial x_4'}\frac{\partial x_4}{\partial u_c}\right|_{u=u_c}^{u=\infty}
+\left.\frac{\partial {\cal L}}{\partial h'}\frac{\partial h}{\partial u_c}\right|_{u=u_c}^{u=\infty} - {\cal L}\Big|_{u=u_c}
= \left(kx_4'+h'\frac{\partial {\cal L}}{\partial h'}-{\cal L}\right)_{u=u_c} \, ,
\eea
where we used the same arguments as explained below Eq.\ (\ref{dOmduc1}). 
We compute
\be
kx_4'+h'\frac{\partial {\cal L}}{\partial h'}-{\cal L} = u^{5/2}\sqrt{1+g_1}\sqrt{1+g_2-\frac{k^2}{u^8f_T}+\frac{(n_IQ)^2}{u^5}}
\left[\frac{g_1}{1+g_1}-\frac{1+g_2-\frac{k^2}{u^8f_T}}{1+g_2-\frac{k^2}{u^8f_T}+\frac{(n_IQ)^2}{u^5}}\right] \, .
\ee
Since $Q(u_c)=0$, we obtain 
\be
\left(kx_4'+h'\frac{\partial {\cal L}}{\partial h'}-{\cal L}\right)_{u=u_c} = -\left.\frac{u^{5/2}}{\sqrt{1+g_1}}\sqrt{1+g_2-\frac{k^2}{u^8f_T}}\right|_{u=u_c} = 0 \, , 
\ee
because $g_1\propto (u-u_c)^{-1}$, and $g_2 = {\rm const}$ at $u=u_c$, i.e., we are already sitting on a stationary point of the free energy with respect to $u_c$ once we fulfill
the equations of motion.

\begin{figure} [t]
\begin{center}
\underline{Deconfined geometry, homogeneous ansatz, $T=0$}

\vspace{0.2cm}
\hbox{\includegraphics[width=0.5\textwidth]{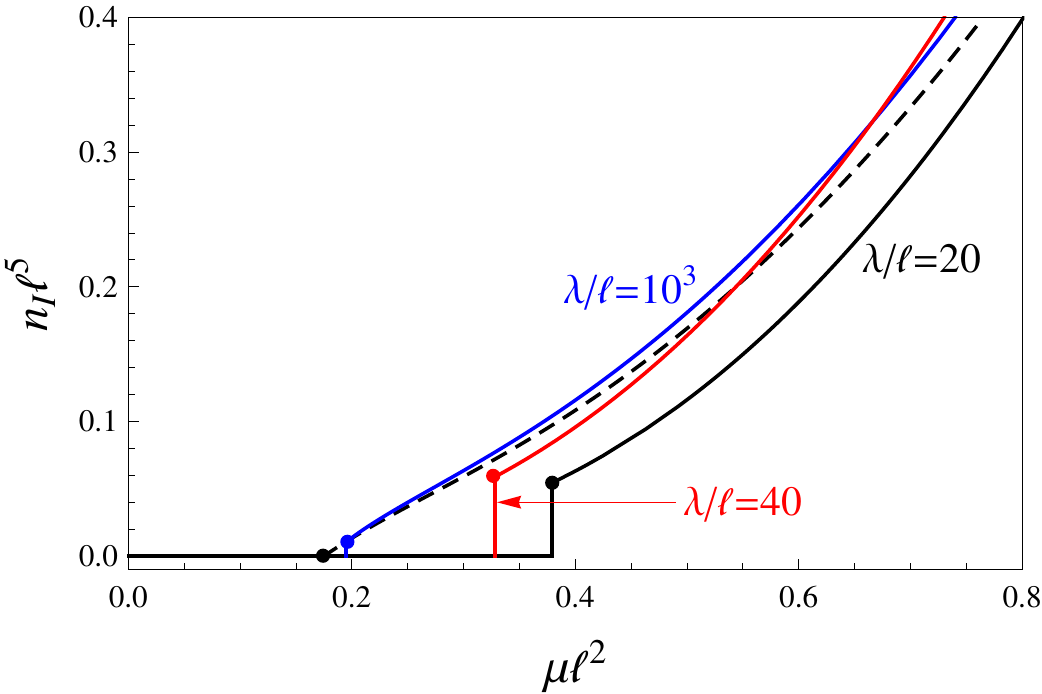}\includegraphics[width=0.5\textwidth]{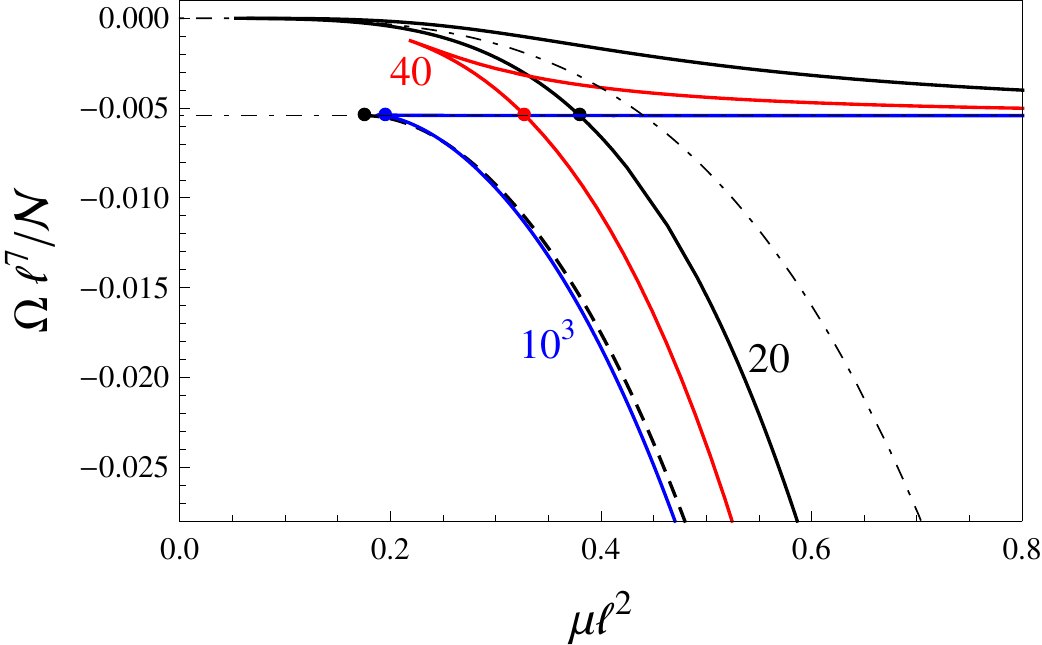}}

\hbox{\includegraphics[width=0.495\textwidth]{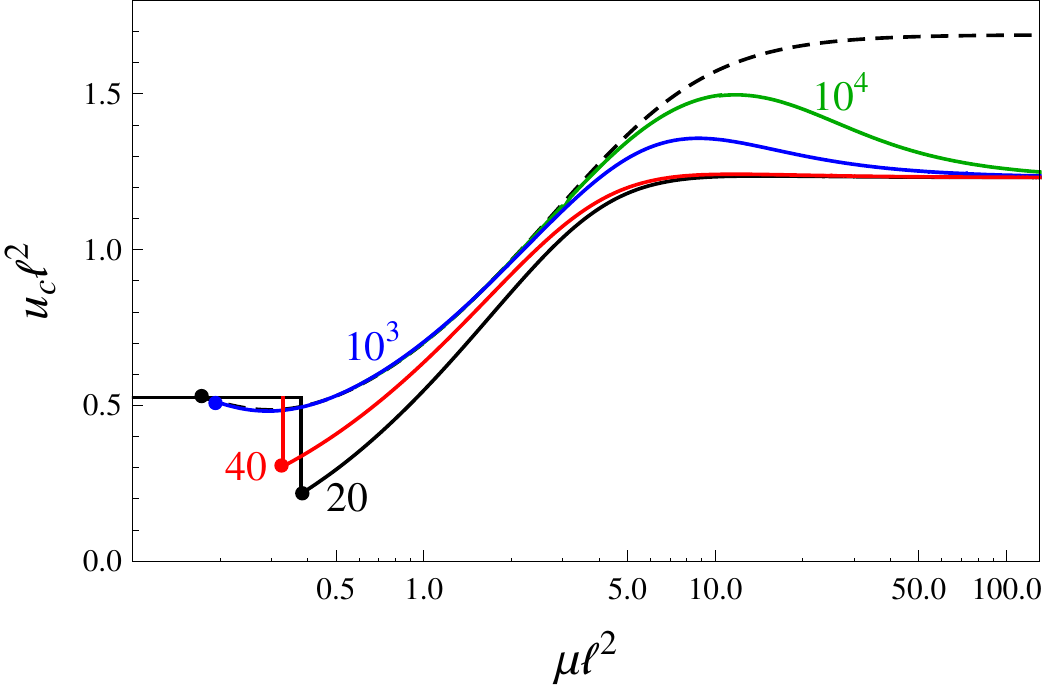}\hspace{0.2cm}\includegraphics[width=0.495\textwidth]{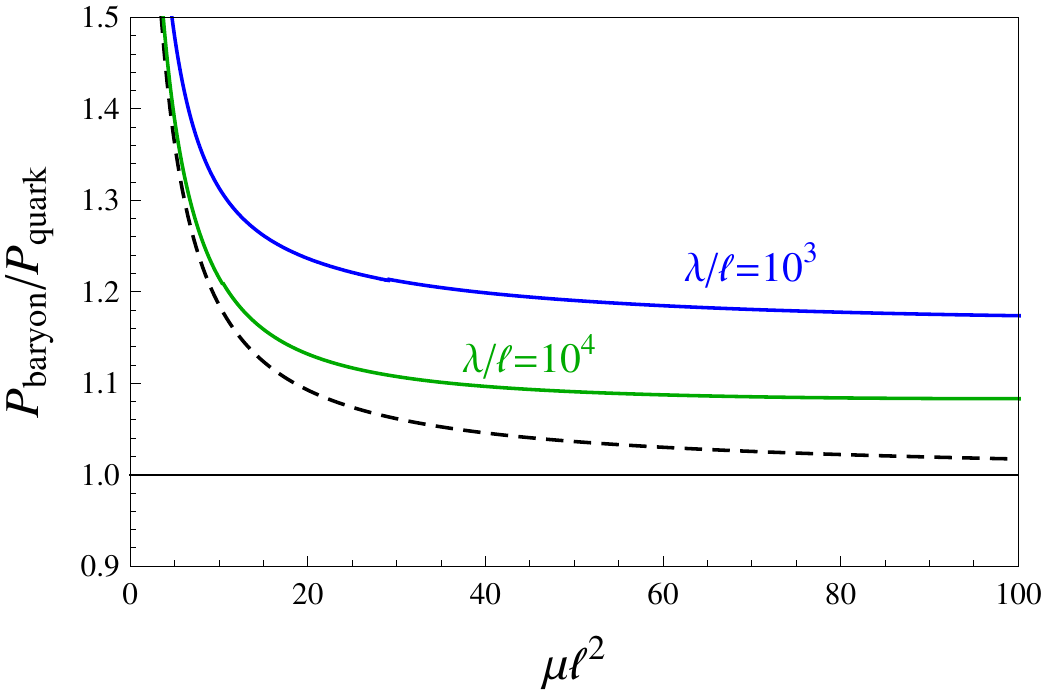}}
\caption{{\it Upper left panel:} zero-temperature baryon density for three different values of the 't Hooft coupling as a function of the chemical 
potential, showing the discontinuity at the first-order transition. In all panels of this figure, the dashed line is the result for pointlike baryons.
{\it Upper right panel:}
corresponding free energy, including the metastable and unstable branches. The thin dashed-dotted lines are the free energies of the mesonic phase (horizontal line)
and the chirally restored phase.
{\it Lower left panel:}  location of the tip of the connected flavor branes on a large, logarithmic $\mu$ scale.  
{\it Lower right panel:} ratio of the pressures of the baryonic and chirally restored phases for two large values of the 't Hooft coupling on a large, linear $\mu$ scale.}       
\label{figdeconf2}
\end{center}
\end{figure}

\begin{figure} [t]
\begin{center}
\includegraphics[width=0.5\textwidth]{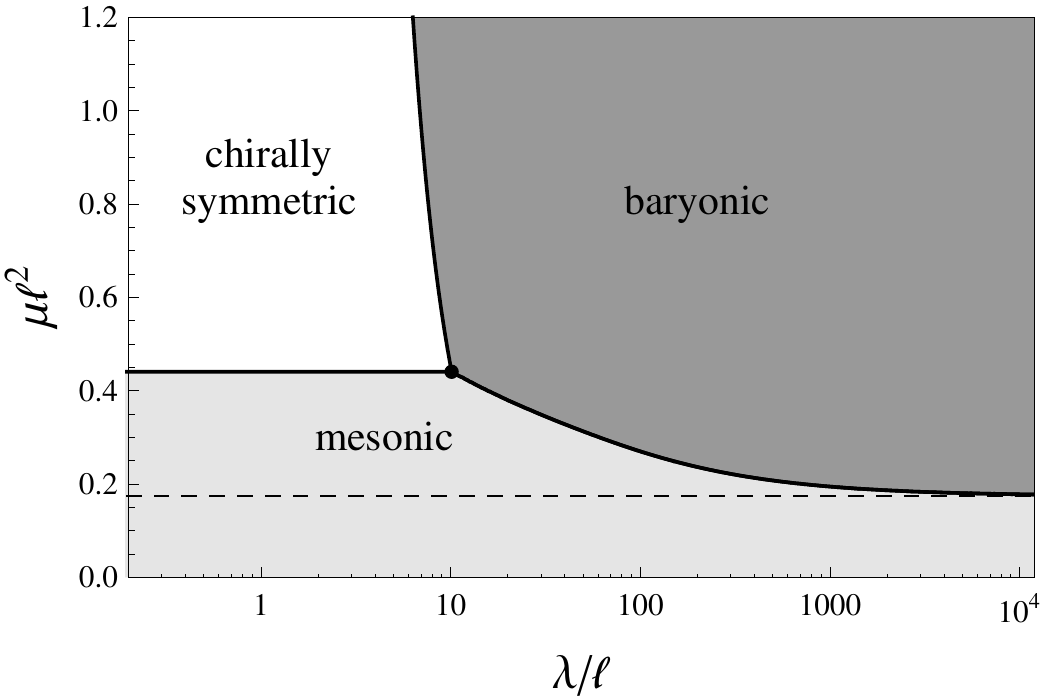}
\caption{Phases in the plane of chemical potential $\mu$ and 't Hooft coupling $\lambda$ at zero temperature for the homogeneous ansatz of the non-abelian gauge 
fields. All solid lines are first-order phase transitions. The thin dashed line marks the (second-order, $\lambda$-independent) baryon onset for the instanton gas. For not too
small values of the 't Hooft coupling, there is a first-order phase transition from the vacuum to baryonic matter, but -- if the coupling is kept fixed -- no 
subsequent transition to chirally restored matter.}
\label{figphaseshom}
\end{center}
\end{figure}

Finally, we need to fulfill the boundary conditions (\ref{boundarya}) and (\ref{fixell}). For a given chemical potential $\mu$, these two boundary conditions, together 
with the equation for $h$ (\ref{eqh1}), determine $a_1$, $n_I$, and $u_c$. Instead of solving this complicated system of equations simultaneously, we can proceed as follows. 
First, we rescale all quantities by appropriate powers of $u_c$ and the asymptotic separation $\ell$, as explained in Sec.\ \ref{sec:numerical}, see Table \ref{table1}. In addition
to the quantities listed in that table, we need $\lambda \sim u_c^{-1/2}$, $a_1\sim u_c^{1/2}$, $h\sim u_c$ and accordingly for $\ell$, using $\ell\sim u_c^{-2}$. Then, $\ell$ is 
completely eliminated from the equations, while $u_c$ only appears in a trivial way in the boundary condition (\ref{fixell}), which thus decouples. 
We now choose a value for (the rescaled)
$n_I$, and determine $a_1$ and $h(u)$ via the shooting method from Eq.\ (\ref{eqh1}). Then, after determining $\hat{a}_0(u_c)$ from Eq.\ (\ref{a0uc2}), we use the 
boundary condition (\ref{boundarya}) in the form
\be
\mu = \int_{u_c}^\infty du\,\hat{a}_0' +\hat{a}_0(u_c) 
\ee
to compute (the rescaled) $\mu$.  [Computing $\mu$ as a function of $n_I$ is advantageous also because it is a single-valued 
function, while $n_I(\mu)$ is two-valued.] The disadvantage of this procedure is that we only work with 
rescaled quantities, and the rescaling with $u_c$ has to be undone to obtain the final results (the rescaling with $\ell$ is trivial because $\ell$ is a 
constant). As a consequence, we have used this procedure for
the zero-temperature phase diagram in Fig.\ \ref{figphaseshom}, but it cannot be used when we wish to work with a fixed 't Hooft coupling $\lambda$. In this case, 
we must not rescale $\lambda$, and the numerics become somewhat more difficult because we now have to simultaneously solve the differential equation for $h$ {\it and} the boundary 
condition (\ref{fixell}) to determine $u_c$. This has been done for the plots in Fig.\ \ref{figdeconf2}.

For the free energy comparison to the mesonic and chirally symmetric phases, we notice that, due to our choice of notation, the free energy has exactly the same form as given in Eq.\ (\ref{OmcalN}), only with different functions $\zeta$, $g_2$, and $Q$.

\begin{figure} [t]
\begin{center}
\underline{Confined geometry, homogeneous ansatz}

\vspace{0.2cm}
\hbox{\includegraphics[width=0.5\textwidth]{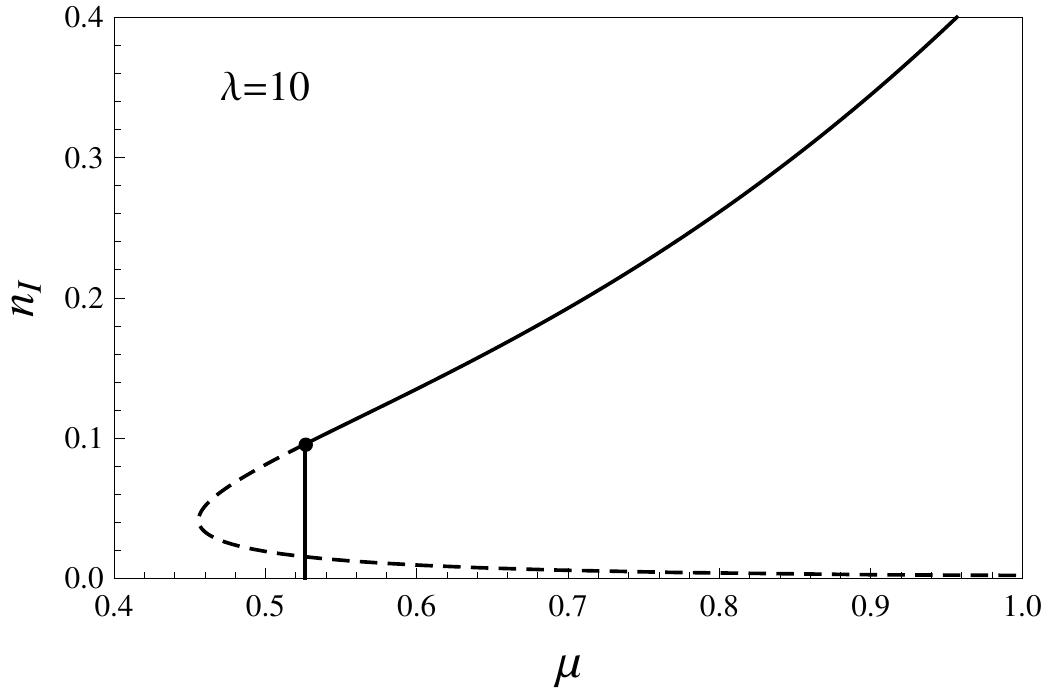}\includegraphics[width=0.5\textwidth]{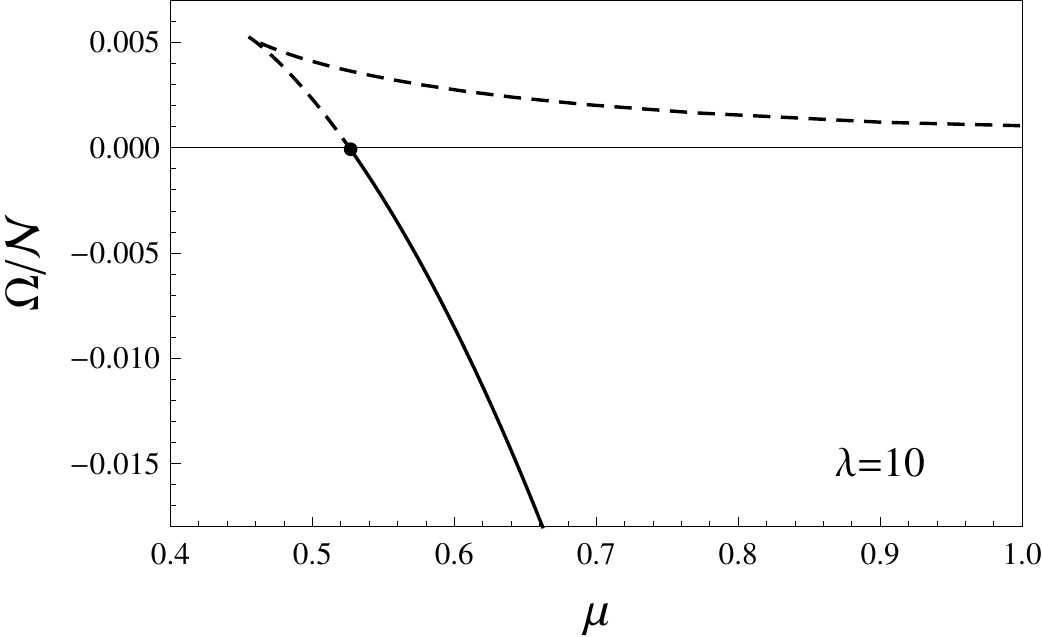}}

\hbox{\includegraphics[width=0.5\textwidth]{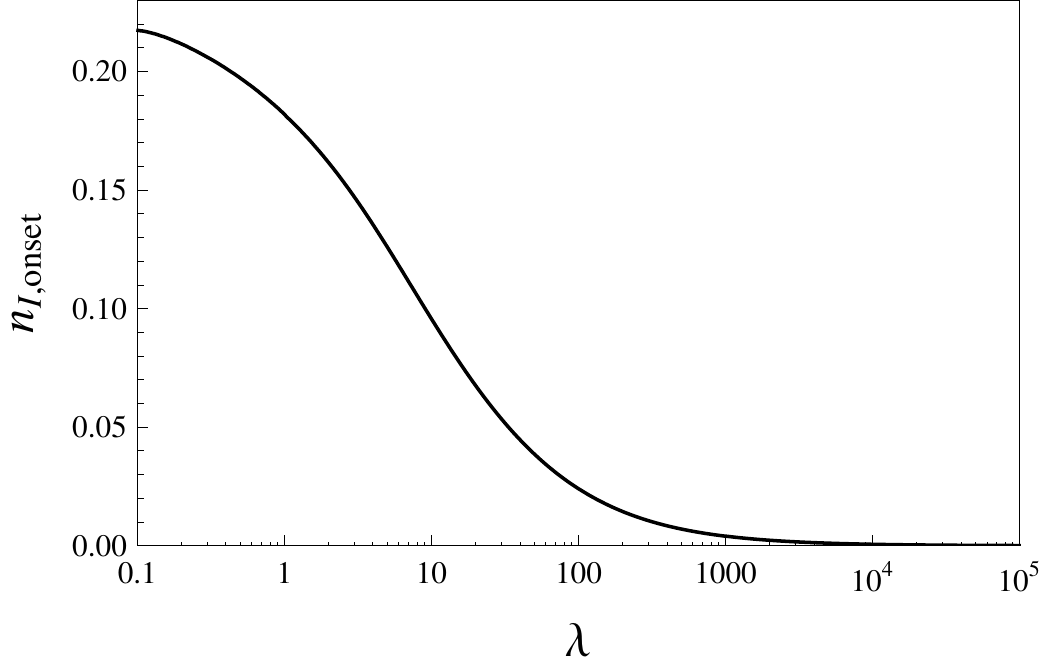}\includegraphics[width=0.5\textwidth]{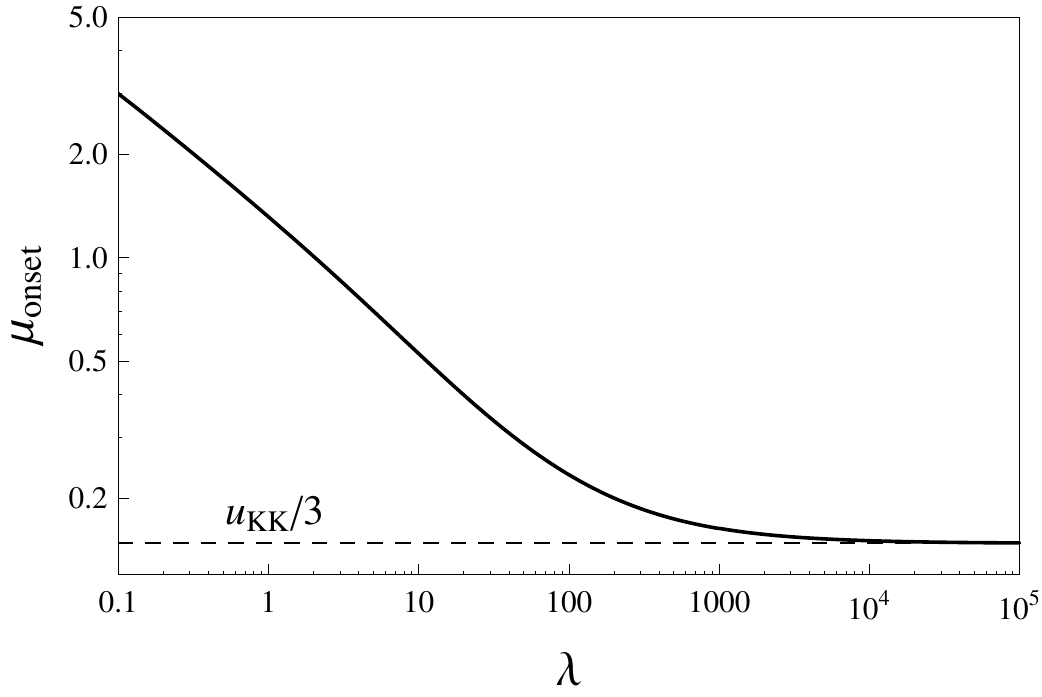}}
\caption{{\it Upper panels:} first-order baryon onset for the confined geometry with maximally separated flavor branes 
within the homogeneous ansatz at a fixed 't Hooft coupling $\lambda=10$. The left panel shows
the baryon density, which is discontinuous at the onset, while the right panel shows the free energy which favors baryonic matter for $\Omega<0$.
In both panels, the dashed continuation of the curve shows the metastable solution (up to the turning point) and the unstable solution.  
{\it Lower panels:}
 baryon density just above and chemical potential at the first-order baryon onset as a function of the 't Hooft coupling $\lambda$; the critical chemical potential approaches 
that of the second-order onset in the pointlike approximation $\mu=u_{\rm KK}/3$ for $\lambda \to \infty$.}       
\label{figconf2}
\end{center}
\end{figure}

\subsection{First-order baryon onset and $\lambda$-$\mu$ phase diagram}
\label{sec:1st}

Again we start our discussion of the numerical results with the confined geometry with maximally separated flavor branes, whose equations for the homogeneous ansatz are 
collected in appendix \ref{app:confhom}. The results are shown in Fig.\ \ref{figconf2}. The two upper panels, where we plot the baryon density and the free energy as a function
of $\mu$ at a fixed value of  $\lambda$, show a first-order baryon onset. This is in accordance with Ref.\ \cite{Rozali:2007rx}, where this observation was already made
for the confined geometry, 
but without a full numerical evaluation. The lower two panels show the dependence on the \mbox{'t Hooft} coupling $\lambda$ of the baryon density just above the onset $n_{I,{\rm onset}}$ and of 
the chemical potential at the onset $\mu_{\rm onset}$. We see that $n_{I,{\rm onset}}$ vanishes for $\lambda\to \infty$, while $\mu_{\rm onset}$
approaches that of the second-order onset of the pointlike approximation and the instanton gas of Sec.\ \ref{sec:inst}. 

In the deconfined geometry we restrict ourselves to zero temperature and present our results in Figs.\ \ref{figdeconf2} and \ref{figphaseshom}. 
In the upper left panel of Fig.\ \ref{figdeconf2} we plot the baryon density as a function of the chemical potential for three values of the 't Hooft coupling, 
$\lambda/\ell = 20,40,10^3$. As for the confined geometry, we obtain two solutions, but only show the stable branch to keep the plot simple. We see that the jump in $n_I$ becomes 
small for large $\lambda$ (although the size of the jump is not a monotonic function of $\lambda$), and the critical chemical potential approaches that of the pointlike
approximation. In the upper right panel we show the free energy of the solutions, in comparison to the mesonic and the chirally restored phases. Here we have included the metastable 
and unstable branches, whose free energy is larger than that of the mesonic phase for all $\mu$. 
In the lower left panel we show the location of the tip of the connected D8- and $\overline{\rm D8}$-branes on a very large $\mu$ scale. We see that 
$u_c$ saturates at a finite value for $\mu\to \infty$, just like for pointlike baryons and in contrast to the instanton gas approximation. The value at infinitely 
large $\lambda$ and $\mu$ depends on the order in which these limits are taken: if $\mu$ is fixed to any arbitrarily large, but finite value, $u_c$ approaches the pointlike result, 
$u_c\simeq 1.69/\ell^2$, for $\lambda\to \infty$; if, on the other hand, $\lambda$ is held fixed, $u_c$ approaches a smaller value for $\mu\to\infty$, namely $u_c\simeq 1.23/\ell^2$. 
Finally, in the lower right panel, we show the ratio of the pressures of the baryonic and chirally restored phases, in analogy to the lower right panel of Fig.\ \ref{figdeconf1}. 
This plot demonstrates that there is no chiral restoration: for any finite value of $\lambda$, the ratio approaches a finite value larger than 1 for $\mu\to \infty$.

We may now compute the critical chemical potential for the baryon onset for all $\lambda$. The resulting zero-temperature phase diagram in the $\lambda$-$\mu$ plane
is shown in Fig.\ \ref{figphaseshom}. As the results of Fig.\ \ref{figdeconf2} have already suggested, we recover the result of the pointlike approximation 
at very large $\lambda$: the baryon onset becomes a weak first-order transition (second order for $\lambda\to \infty$), and we even reproduce asymptotically -- 
within the numerical accuracy and performing the calculation up to $\lambda \simeq 1.4\times 10^4$ -- the numerical value for the critical chemical potential, $\mu\simeq 0.17/\ell^2$. 
Going to smaller values of $\lambda$, the critical chemical potential increases and the system behaves as just discussed and shown in the 
other three panels. For sufficiently low values of $\lambda$, the situation differs qualitatively. In this case, the mesonic phase is not superseded by baryonic matter, but
a chiral phase transition to quark matter occurs, and only at very large $\mu$ baryonic matter becomes favored, resulting in another chiral phase transition.
The topology of the phase structure in the $\lambda$-$\mu$ plane gives rise to a tricritical point at $(\lambda,\mu)\simeq (10.1\ell,0.44/\ell^2)$, where all three phases coexist.  
            
The main conclusion from this phase diagram is that, at fixed 't Hooft coupling, once baryonic matter is created in a realistic first-order phase transition,
it never wants to disappear again, i.e., the baryon onset is never superseded by chiral restoration at high densities, as one would expect from QCD. 
In this sense, the homogeneous ansatz behaves exactly opposite to the instanton gas, where the onset was an unrealistic second-order phase transition, but there
was chiral restoration at high densities. 

Interestingly, the phase diagram of Fig.\ \ref{figphaseshom} shows that if the 't Hooft coupling were allowed to vary as a function of $\mu$, 
it would be possible to move from the vacuum through a first-order onset into the baryonic phase and then via a chiral phase transition to the quark matter phase, as expected 
from QCD. To this end, the 't Hooft coupling $\lambda = g^2 N_c$ would have to decrease with $\mu$, in accordance with the running of the QCD coupling $g$. It is thus 
tempting to speculate what the trajectory of the running coupling would be in the phase diagram of Fig.\ \ref{figphaseshom}. 
Suppose we reach the quark matter phase with a running coupling. 
Then, we know that at some sufficiently large $\mu$ the QCD coupling runs logarithmically, $g^2\propto 1/\ln\mu$. The transition line between the 
quark matter and baryonic phases in Fig.\ \ref{figphaseshom}, however, 
decreases with $\lambda$ through a power law (the numerics suggest $\mu\propto \lambda^{-5/2}$, but computing this phase transition line numerically becomes 
increasingly difficult for large $\mu$). As a consequence, 
it seems the system necessarily has to re-enter the baryonic phase at large $\mu$. But, we need to keep in mind that the current model cannot be expected to be valid in the 
asymptotically dense regime where the system becomes weakly coupled. Therefore, it is perfectly conceivable that the (strong) coupling runs through the three phases of 
Fig.\ \ref{figphaseshom} in the ``right order'' before we anyway have to stop trusting our calculation. The chiral phase transition is expected to 
occur at moderate, not asymptotically large, densities.  
Of course, in the present approximation, there is no prediction for how $\lambda$ might run, and thus here we do not make any attempts to model such a running. 
Possibly this question can be addressed by taking into account the backreaction of the flavor branes on the background geometry, which appears to result in a running coupling,
although different from QCD \cite{Bigazzi:2014qsa}.

\section{Conclusions}
\label{sec:conclusions}

We have studied cold and dense baryonic matter in the Sakai-Sugimoto model and have addressed the question whether the model can account for a first-order baryon onset {\it and} 
a zero-temperature chiral phase transition from baryonic to quark matter. To this end, we have focused on the deconfined geometry and the decompactified limit, where the 
separation of the flavor branes at the holographic boundary is small compared to the radius of the compactified extra dimension of the model.  To our knowledge, this is the 
first time that holographic baryonic matter beyond the simple pointlike approximation has been considered in this setup. 
As a reference calculation and for a comparison to existing results in the literature we have also considered the confined geometry with maximally separated flavor branes. 

In our description of baryonic matter we followed two different approaches. Firstly, we have considered an instanton gas \cite{Ghoroku:2012am}, which makes
use of the BPST instanton solution that has been employed to study vacuum properties of baryons in the Sakai-Sugimoto model \cite{Hata:2007mb}. This approach can be considered
as a generalization of the pointlike approximation for baryonic matter \cite{Bergman:2007wp}. We have recovered the pointlike approximation in 
our calculation for the case of small baryon densities. It has turned out that the baryon density is allowed to become arbitrarily small, resulting in a second-order phase 
transition from the 
vacuum to baryonic matter. At large densities, however, the instanton gas differs qualitatively from pointlike baryons. Most notably, the medium-dependent baryon mass increases without 
bound, such that baryonic matter becomes energetically more and more costly. As a consequence, and in contrast to the pointlike approximation, chiral symmetry is restored 
at zero temperature at (extremely) large densities. 

Secondly, we have employed a homogeneous ansatz for the non-abelian gauge fields, which only depends on the holographic direction \cite{Rozali:2007rx}. 
This approach, in contrast to the instanton gas approximation, allows for a nontrivial dependence on the 't Hooft coupling $\lambda$ and thus 
for going away from the infinite coupling limit (at least in an extrapolating sense since we do not compute finite-$\lambda$ corrections systematically). 
This is important because only for finite values of the 't Hooft coupling do we see a first-order baryon onset, as expected from QCD.
If $\lambda$ is kept fixed and not too small, we have found that after baryonic matter is created in a first-order phase transition, it remains favored for all densities, i.e., 
there is no chiral restoration. A sequence of phases with increasing chemical potential that is expected from QCD (vacuum -- nuclear matter -- quark matter) is only 
possible if $\lambda$ were to decrease with increasing chemical potential. Although this is an intriguing observation since this is exactly what is expected from the 
QCD coupling, the present approximation does not include any running of $\lambda$. The main conclusion from our two different approaches is thus that neither one shows a first-order
baryon onset {\it and} chiral restoration at high density, but we see chiral restoration in the instanton gas approximation and a first-order onset in the homogeneous approach.

In order to interpret this conclusion we need to keep in mind that we have employed several approximations and extrapolations. We have used a simple prescription for the 
non-abelian DBI action and neglected any backreaction of the flavor branes on the metric. Therefore, our results for very large densities have to be taken with some care. 
We have used the simple flat-space, $SO(4)$ symmetric BPST instanton solution. In going from a single instanton profile to a gas of instantons, we have employed a simple spatial 
averaging in order to obtain homogeneous baryonic matter and have placed all instantons at the tip of the connected flavor branes, i.e., we have not determined their location in the bulk 
dynamically. All these deficiencies of our calculation can be improved in a systematic way, and thus we believe that our observations will help to move closer towards a holographic picture
of real-world dense matter. For example, our observation of the first-order phase transition in the homogeneous approach is a clear indication that some of the physics away 
from the $\lambda=\infty$ limit must be included, and we are currently working on extending the instanton gas approach in this direction.

We have mainly focused on two crucial properties of dense matter, a first-order baryon onset and chiral restoration at large densities. Of course, 
fulfilling these necessary properties does not guarantee that we are quantitatively close to real-world QCD. Indeed, as we have argued, there are reasons to expect that, at best, 
the Sakai-Sugimoto model in the decompactified version can only be a rough guide to high-density QCD. Nevertheless, the lack of first-principle calculations and 
of {\it any} model for cold matter that is reasonably applicable over a wide density regime would make even a semi-realistic strong-coupling model very valuable.
The Sakai-Sugimoto model in the limit considered here has only three free parameters, the 't Hooft coupling, the Kaluza-Klein mass, and the asymptotic separation of the 
flavor branes. At this point, we have not attempted to fit them to any physical properties. After the model has been improved along the lines just indicated, 
it will be very interesting to fit them for instance to nuclear ground state properties and use the model for quantitative predictions, e.g.,  for the equation of state 
in the context of compact stars.

\begin{acknowledgments}
We would like to thank Frederic Br\"{u}nner, 
Florian Preis, Anton Rebhan, and Stefan Stricker for valuable comments and discussions. SWL and QW are supported partially by the Major State Basic Research Development Program in China under the grant No.~2015CB856902 and the National Natural
Science Foundation of China under the grant No.~11125524. AS acknowledges support from the Austrian science 
foundation FWF under project no.\ P26328 and by the {\mbox NewCompStar} network, COST Action MP1304.  We thank the Kavli Institute for Theoretical Physics China (KITPC), Beijing, for its kind hospitality during the program "sQGP and extreme QCD", where this work was finalized. 
\end{acknowledgments}

\appendix

\section{Confined geometry}
\label{app:conf}

All derivations in the main text are done in the deconfined geometry. In this appendix we discuss the confined geometry which, in our context, mainly serves as a 
warm-up exercise for the numerical calculations of the deconfined geometry. The reason is that it allows for a very simple scenario where the embedding of the 
flavor branes in the background geometry is trivial. Most of the arguments are completely analogous (in many instances simpler) as in the deconfined geometry.

The induced metric on the flavor branes is
\begin{subequations}
\bea
ds^2_{\rm D8} = \left(\frac{U}{R}\right)^{3/2}(d\tau^2 + \delta_{ij} dX^i dX^j) + \left(\frac{R}{U}\right)^{3/2}\left\{\left[\frac{1}{f}+f\left(\frac{U}{R}
\right)^3(\partial_UX_4)^2\right] dU^2 +U^2d\Omega_4^2\right\} \, , 
\eea
\end{subequations}
where
\be
f \equiv  1-\frac{U_{\rm KK}^3}{U^3} \, , 
\ee
and $U_{\rm KK}$ is related to the Kaluza-Klein mass through
\be \label{MKK}
M_{\rm KK}=\frac{3}{2}\frac{U_{\rm KK}^{1/2}}{R^{3/2}} \, .
\ee

\subsection{Baryonic matter from an instanton gas}
\label{app:confinst}

As for the deconfined geometry, we first expand the action for small non-abelian field strengths, use this expansion to introduce the instanton gas with the help of the 
instanton solution of the YM action, and then consider the action to all orders in the non-abelian field strengths. For the expansion we need 
\bea \label{expconf1}
\sqrt{{\rm det}(g+2\pi\alpha'\hat{F})} = \frac{U^4}{\sqrt{f}}\left(\frac{R}{U}\right)^{3/4}\sqrt{1+u^3 f^2 x_4'^2-f\hat{a}_0'^2} \, , \\[2ex]
\label{expconf2}
\Tr\left[(g+2\pi\alpha'\hat{F})^{-1} F\right]^2 = -\left(\frac{R}{U}\right)^{3}\Tr[F_{ij}^2]-\frac{2f\Tr[F_{iU}^2]}{1+u^3f^2x_4'^2-f\hat{a}_0'^2} \, ,
\eea
which gives the DBI action
\be \label{SDBIconf}
S_{\rm DBI} = \frac{2T_8V_4R^{3/2}}{g_s} \int d^4X  \int_{U_{\rm KK}}^\infty dU \, \frac{U^{5/2}}{\sqrt{f}} 
\sqrt{1+u^3 f^2 x_4'^2-f\hat{a}_0'^2} \left[1+\frac{R^3(2\pi\alpha')^2}{4U^3}\Tr[F_{ij}^2]
+\frac{f}{2}\frac{(2\pi\alpha')^2\Tr[F_{iU}^2]}{1+u^3f^2x_4'^2-f\hat{a}_0'^2}\right] \, . 
\ee
In order to discuss the instanton solution, we need to consider the YM action, which is immediately obtained from this expansion. We consider the case of maximally 
separated branes, such that the embedding of the flavor branes is trivial, $x_4'=0$ (for a generalization to the non-antipodal case, see Ref.\ \cite{Seki:2008mu}).
We also set the chemical potential to zero, i.e., $\hat{a}_0(\infty)=0$.
Nevertheless, due to the CS term, $\hat{a}_0(u)$ acquires a nontrivial profile even in that case \cite{Hata:2007mb}, which affects the instanton solution. For now, we also 
ignore the CS contribution, such that we can set $\hat{a}_0(u)=0$. Then, dropping the term that is independent of the gauge fields,  
we obtain the YM action \cite{Sakai:2004cn,Sakai:2005yt,Hata:2007mb},
\be
S_{\rm YM} = \frac{\lambda N_c}{216 \pi^3} \int d^4X  
\int_{-\infty}^\infty dZ\,\left[\frac{h(Z)}{2}\Tr[F_{ij}^2]+\frac{M_{\rm KK}^2}{U_{\rm KK}^2}k(Z)\Tr[F_{iZ}^2]\right] \, , 
\ee
where we have used Eqs.\ (\ref{constants}) and (\ref{MKK}), the variable $Z$ from Eq.\ (\ref{Z}) with $U_c$ replaced by $U_{\rm KK}$
(we now integrate over both halves of the connected flavor branes), 
and abbreviated $k(Z) \equiv U_{\rm KK}^3 + Z^2 U_{\rm KK}$, $h(Z)\equiv (U_{\rm KK}^3+ Z^2U_{\rm KK})^{-1/3}$. 

In flat space, setting $k(Z) = U_{\rm KK}^3$ and $h(Z) = U_{\rm KK}^{-1}$, the YM equations of motion are solved by the instanton solution \cite{Hata:2007mb}
\be \label{AzAiinst}
A_{Z}(\vec{X},Z) = -i\phi \psi\partial_{Z}\psi^{-1} \, , \qquad A_i(\vec{X},Z) = -i\phi \psi\partial_i\psi^{-1} \, , 
\ee
with 
\be
\phi(\vec{X},Z) = \frac{\xi^2}{\xi^2+(\rho/\gamma)^2} \, , \qquad \psi(\vec{X},Z) = \frac{Z/\gamma-i\vec{X}\cdot \vec{\sigma}}{\xi} \,  , \qquad \xi^2\equiv (\vec{X}-\vec{X}_0)^2
+\frac{(Z-Z_0)^2}{\gamma^2}  \, , 
\ee
where we have abbreviated 
\be \label{rho0}
\gamma \equiv M_{\rm KK} U_{\rm KK} = \frac{3U_{\rm KK}^{3/2}}{2R^{3/2}} \, .
\ee
The reason for the appearance of this factor, which rescales $Z$ relative to $\vec{X}$, is the relative factor between the $F_{ij}^2$ and $F_{iZ}^2$ terms in the YM action 
(we have taken the flat-space limit by dropping the $Z$ dependence, but not setting the constant prefactors to one). 
Since the instanton width will later be determined dynamically, we are free to denote it by 
$\rho/\gamma$. This will turn out to be convenient when we take the spatial average of the instanton gas, see Eq.\ (\ref{gas}), after which $\gamma$ simply becomes a prefactor of the 
instanton profile.   

With $F_{\mu\nu} = \partial_\mu A_\nu - \partial_\nu A_\mu + i[A_\mu,A_\nu]$, the resulting field strengths are
\be \label{FijFiZ}
F_{ij} = \epsilon_{ija}\sigma_a\frac{2(\rho/\gamma)^2}{[\xi^2+(\rho/\gamma)^2]^2} \,, \qquad F_{iZ} = -\frac{\sigma_i}{\gamma} \frac{2(\rho/\gamma)^2}{[\xi^2+(\rho/\gamma)^2]^2} \, . 
\ee
The generalization of this single instanton to a gas of instantons is explained in the main text for the deconfined geometry, see Sec.\ \ref{sec:instantons}.
In our approximation, this generalization results in the field strengths (\ref{Fus}). Completely analogously, we obtain for the confined geometry
\bea \label{Fusconf} 
\Tr[F_{iU}^2] \;\to\; \frac{u^{1/2}}{3u_{\rm KK}^2\sqrt{f}} \frac{n_Iq(u)}{(2\pi\alpha')^2} \, , \quad 
\frac{\Tr[F_{ij}^2]}{(M_{\rm KK}R)^6} \;\to\;  \frac{2u_{\rm KK}^2\sqrt{f}}{3u^{1/2}} \frac{n_Iq(u)}{(2\pi\alpha')^2} \, ,\quad 
\frac{\Tr[F_{ij}F_{kU}]\epsilon_{ijk}}{(M_{\rm KK}R)^3} \;\to\; - \frac{2}{3} \frac{n_Iq(u)}{(2\pi\alpha')^2} \,, 
\eea
with
\be \label{qconf}
q(u) = \frac{9u^{1/2}}{4\sqrt{f}}\frac{(\rho^2u_{\rm KK})^2}{(u^3-u_{\rm KK}^3+\rho^2u_{\rm KK})^{5/2}} \, . 
\ee
With $u_{\rm KK}$ defined in Table \ref{table0} and using Eq.\ (\ref{MKK}), we have
\be
u_{\rm KK} = \frac{4}{9} \, .
\ee
As in the deconfined case, our actual calculation is performed by keeping all orders in the field strengths. For the abelian case, this yields [compare with the corresponding 
expression of the deconfined geometry (\ref{DBIdeconf})]
\bea \label{DBIconf}
\sqrt{{\rm det}(g+2\pi\alpha'\hat{F})} 
&=& \frac{U^4}{\sqrt{f}}\left(\frac{R}{U}\right)^{3/4}\left\{(2\pi\alpha')^2f\hat{F}_{iU}^2+\left[1+u^3 f^2 x_4'^2
+f\hat{a}_0'^2\right]\left[1+\left(\frac{R}{U}\right)^3\frac{(2\pi\alpha')^2\hat{F}_{ij}^2}{2}\right] \right.\non[2ex]
&&\hspace{2cm}\left. + \left(\frac{R}{U}\right)^3\frac{(2\pi\alpha')^4f 
(\hat{F}_{ij}\hat{F}_{kU}\epsilon_{ijk})^2}{4}\right\}^{1/2} \, .
\eea
We now generalize this expression to the non-abelian case, following the prescription explained at the beginning of Sec.\ \ref{sec:allorders}. With the help 
of Eq.\ (\ref{Fusconf}), working with maximally separated flavor branes, $x_4'=0$, and including the CS term, we arrive at the Lagrangian 
\be \label{Su}
{\cal L}= \frac{u^{5/2}}{\sqrt{f}}\sqrt{(1+g_1-f \hat{a}_0'^2)(1+ g_2)} - n_I \hat{a}_0(u)q(u) \, ,
\ee
where 
\be
g_1(u) \equiv \frac{u^{1/2}\sqrt{f(u)}}{3u_{\rm KK}^2}n_Iq(u) \, , \qquad g_2(u) \equiv \frac{u_{\rm KK}^2\sqrt{f(u)}}{3u^{7/2}}n_Iq(u) \, ,
\ee
with $q(u)$ from Eq.\ (\ref{qconf}). The solution of the equation of motion for $\hat{a}_0$ is given by 
\be
a_0'^2 = \frac{(n_IQ)^2}{u^5f}\frac{1+g_1}{1+g_2+\frac{(n_IQ)^2}{u^5}} \, ,
\ee
where 
\be
Q(u) = \frac{u^{3/2}\sqrt{f}}{2}\frac{3\rho^2u_{\rm KK}+2(u^3-u_{\rm KK}^3)}{(u^3-u_{\rm KK}^3+\rho^2u_{\rm KK})^{3/2}} \, .
\ee
Following the arguments in Sec.\ \ref{sec:allorders}, we minimize the free energy with respect to $n_I$ and $\rho$, which leads to the 
coupled equations
\begin{subequations} \label{miniconf}
\bea
0 &=&  \int_{u_{\rm KK}}^\infty du\, \left[\frac{u^{5/2}}{2\sqrt{f}}\left(\frac{\partial g_1}{\partial n_I}\zeta^{-1}
+\frac{\partial g_2}{\partial n_I}\zeta \right) + \hat{a}_0'Q\right]-\mu \, , \label{miniconf1}\\[2ex]
0 &=&  \int_{u_{\rm KK}}^\infty du\, \left[\frac{u^{5/2}}{2\sqrt{f}}\left(\frac{\partial g_1}{\partial \rho}\zeta^{-1}
+\frac{\partial g_2}{\partial \rho}\zeta \right) + n_I \hat{a}_0'\frac{\partial Q}{\partial \rho}\right] \, , \label{miniconf2} 
\eea
\end{subequations}
where
\be
\zeta\equiv \sqrt{\frac{1+g_1-f\hat{a}_0'^2}{1+g_2}} = \frac{u^{5/2}\sqrt{f}\hat{a}_0'}{n_IQ} = \frac{\sqrt{1+g_1}}{\sqrt{1+g_2+\frac{(n_IQ)^2}{u^5}}} \, , 
\ee
and we can write the free energy as
\be
\frac{\Omega}{\cal N} = \int_{u_{\rm KK}}^\infty du\,\frac{u^{5/2}}{\sqrt{f}}\sqrt{1+g_1}\sqrt{1+g_2+\frac{(n_IQ)^2}{u^5}} - \mu n_I \, .
\ee

\subsection{Baryonic matter from a homogeneous ansatz}
\label{app:confhom}

In the confined geometry with maximally separated flavor branes, the homogeneous ansatz for the gauge fields (\ref{ansatzhom}) yields, in analogy to 
Eq.\ (\ref{Shom}),  the action
\be \label{ShomConf}
S = {\cal N} \int_{u_{\rm KK}}^\infty du\,\left[\frac{u^{5/2}}{\sqrt{f}}\sqrt{(1+g_1-f\hat{a}_0'^2)(1+g_2)}  - \frac{9\lambda_0}{2}\hat{a}_0h^2h'\right] \, ,
\ee
where
\be
g_1(u) \equiv \frac{3f(u)h'(u)^2}{2} \, , \qquad g_2(u) \equiv \frac{3\lambda_0^2h(u)^4}{2u^3} \, .
\ee
We now proceed completely analogously to the deconfined geometry, as explained in Sec.\ \ref{sec:ancalc}: the baryon density is related to the boundary value of $h$ in the bulk,
$n_I = -3\lambda_0/2\,h(u_{\rm KK})^3$, and the equation of motion for $\hat{a}_0$ in integrated form is  
\be
\frac{u^{5/2}\sqrt{f}\hat{a}_0'\sqrt{1+g_2}}{\sqrt{1+g_1-f\hat{a}_0'^2}} = n_IQ \, , \qquad Q(u) \equiv 1+\frac{3\lambda_0}{2n_I}h(u)^3 \, ,
\ee
such that 
\be 
\hat{a}_0'^2 = \frac{(n_IQ)^2}{u^5f}\frac{1+g_1}{1+g_2+\frac{(n_IQ)^2}{u^5}} \, .
\ee
With the abbreviation 
\be
\zeta \equiv \frac{\sqrt{1+g_1-f\hat{a}_0'^2}}{\sqrt{1+g_2}} = \frac{\sqrt{1+g_1}}{\sqrt{1+g_2+\frac{(n_IQ)^2}{u^5}}} 
\ee
the equation of motion for $h$ becomes
\be 
\partial_u\left(\frac{u^{5/2}\zeta^{-1}}{\sqrt{f}}\frac{\partial g_1}{\partial h'}\right)-\frac{9\lambda_0h^2\zeta n_I Q}{u^{5/2}\sqrt{f}} = \frac{u^{5/2}\zeta}{\sqrt{f}}
\frac{\partial g_2}{\partial h} \, ,
\ee
and we find that $h$ behaves around $u=u_{\rm KK}$ as given in Eq.\ (\ref{hexp}) for the deconfined geometry. The minimization with respect to $n_I$ yields
\be \label{a0uKKhom}
\hat{a}_0(u_{\rm KK}) = \frac{u_{\rm KK}}{3}\frac{\sqrt{1+\frac{2u_{\rm KK}^3}{3\lambda_0^2h(u_{\rm KK})^4}}}{\sqrt{1+\frac{8u_{\rm KK}}{9a_1^2}}} \, ,
\ee
and the free energy becomes
\be
\frac{\Omega}{\cal N} =  \int_{u_{\rm KK}}^\infty du\,\frac{u^{5/2}}{\sqrt{f}}\sqrt{1+g_1}\sqrt{1+g_2+\frac{(n_IQ)^2}{u^5}} - \mu n_I \, .
\ee
[We have checked that minimizing this expression numerically with respect to $n_I$ gives the same result as using the simpler (\ref{a0uKKhom}).]

\section{Pointlike baryons, mesonic phase, and chirally restored phase}
\label{app:pointlike}

In this appendix we discuss the baryonic phase in the approximation of pointlike baryons as well as the mesonic and chirally restored phases, which are needed 
for the discussion of the phase structure in the main text. Most of the discussion in this appendix concerns the deconfined geometry, but we also 
briefly discuss the (much simpler) results for the baryonic and mesonic phases of the confined geometry with maximally separated flavor branes. The chirally 
restored phase exists only in the deconfined geometry.

\subsection{Pointlike baryons}
\label{sec:pointlike}

\subsubsection{Deconfined geometry}

As argued in the main text, we recover the Lagrangian for pointlike baryons (\ref{Lpoint}) in the limit of vanishing baryon width $\rho\to 0$, 
\be 
{\cal L} = {\cal L}_0 + n_I\left[\frac{u}{3}\sqrt{f_T(u)} - \hat{a}_0(u)\right]\delta(u-u_c) \, ,
\ee
with ${\cal L}_0$ from Eq.\ (\ref{L0}).
This Lagrangian was the starting point in Ref.\ \cite{Bergman:2007wp}, and here we recapitulate the calculation of that reference in a concise way 
(a very similar calculation can be found in Ref.\ \cite{Preis:2011sp}, where a background magnetic field was included). 
The equations of motion in integrated form are 
\begin{subequations}
\bea
\frac{u^{5/2}\hat{a}_0'}{\sqrt{1+u^3f_Tx_4'^2-\hat{a}_0'^2}} &=& n_I \, , \\[2ex]
\frac{u^{5/2}u^3f_Tx_4'}{\sqrt{1+u^3f_Tx_4'^2-\hat{a}_0'^2}} &=& k \, , 
\eea
\end{subequations}
with an integration constant $k$ that has to be determined in the following. These equations are easily solved for $\hat{a}_0'$ and $x_4'$, 
\be\label{A0pX4p}
\hat{a}_0'(u) = \frac{n_I u^{3/2}\sqrt{f_T(u)}}{\sqrt{g(u)}}  \, , \quad x_4'(u) = \frac{k}{u^{3/2}\sqrt{f_T(u)}\sqrt{g(u)}} \, , 
\ee
where we have abbreviated
\be
g(u) \equiv (u^8+u^3n_I^2)f_T(u)-k^2 \, .
\ee
Consequently, $\hat{a}_0$ and $x_4$ are 
\be
\hat{a}_0(u) = n_I \int_{u_c}^u dv\,\frac{v^{3/2}\sqrt{f_T(v)}}{\sqrt{g(v)}} + \hat{a}_0(u_c) \, , \qquad x_4(u) = k \int_{u_c}^u \frac{dv}{v^{3/2}\sqrt{f_T(v)}\sqrt{g(v)}} \, , 
\ee
where we have set $x_4(u_c)=0$. 

As explained in the main text for finite-size baryons, we need to minimize the free energy with respect to $n_I$ and $u_c$. 
From the minimization with respect to $n_I$ we obtain 
\be \label{a0uc}
\hat{a}_0(u_c) = \frac{u_c}{3} \sqrt{f_T(u_c)} \, .
\ee
As a consequence, the on-shell action is solely determined by ${\cal L}_0$. The minimization with respect to $u_c$ yields 
[using Eq.\ (\ref{a0uc})] 
\be
\frac{n_I}{3}\frac{1+f_T(u_c)}{2} = \frac{\sqrt{g(u_c)}}{u_c^{3/2}} \, . 
\ee
This relation can be used to determine $k$ as a function of $u_c$ and $n_I$,
\be \label{kpointlike}
k^2 = (u_c^8+u_c^3n_I^2)f_T(u_c)-u_c^3\left(\frac{n_I}{3}\right)^2\left[\frac{1+f_T(u_c)}{2}\right]^2 \, .
\ee
By inserting this result into the expression for $x_4'(u)$ from Eq.\ (\ref{A0pX4p}), we see that $x_4'(u)$ assumes a finite value at $u=u_c$ for all nonzero $n_I$. Therefore, 
the embedding of the connected flavor branes acquires a cusp, in contrast to the instanton gas approximation, as shown in the main text.
    
The quantities $u_c$ and $n_I$ are then determined by the coupled system of equations,
\be
\frac{\ell}{2} = k\int_{u_c}^\infty \frac{du}{u^{3/2}\sqrt{f_T(u)}\sqrt{g(u)}} \, , \qquad 
\mu-\hat{a}_0(u_c) = n_I\int_{u_c}^\infty du\, \frac{u^{3/2}\sqrt{f_T(u)}}{\sqrt{g(u)}} \, ,
\ee
where we have imposed the boundary condition $\mu = \hat{a}_0(\infty)$. 
Since the right-hand side of the second equation is larger than zero, there can only be solutions for $\mu>\hat{a}_0(u_c)$. In the limit $n_I\to 0$, $\mu$ assumes this minimum value, 
i.e., with Eq.\ (\ref{a0uc}), the critical chemical potential for the second-order onset of baryonic matter is
\be \label{muc}
\mu_{\rm onset} = \frac{u_c}{3} \sqrt{f_T(u_c)} \, , 
\ee
where $u_c$ at $n_I\to 0$ is determined from 
\be \label{uconset}
\frac{\ell}{2} = \int_{u_c}^\infty du\, \frac{u_c^4\sqrt{f_T(u_c)}}{u^{3/2}\sqrt{f_T(u)}\sqrt{u^8f_T(u)-u_c^8f_T(u_c)}} \, .
\ee
At zero temperature, this can be evaluated analytically, and we find
\be
\mu_{\rm onset}(T=0) = \frac{16\pi}{3\ell^2}\left[\frac{7\Gamma\left(\frac{31}{16}\right)\tan\frac{\pi}{16}}{15\Gamma\left(\frac{23}{16}\right)}\right]^2 \simeq \frac{0.17495}{\ell^2} \, .
\ee
By solving Eq.\ (\ref{uconset}) numerically we can compute the baryon onset for all temperatures. The result is the thick dashed line in Fig.\ \ref{figpointlike}.
For the free energy comparison with the mesonic and chirally restored phases we need the free energy 
\be \label{Ompoint}
\Omega_{\rm pointlike} =  {\cal N} \int_{u_c}^\infty du\,u^5\frac{u^{3/2}\sqrt{f_T(u)}}{\sqrt{g(u)}} \, .
\ee
As mentioned in the main text below Eq.\ (\ref{defzeta0}), a divergent vacuum contribution has to be subtracted, as for all free energies in this paper. 
With this simple renormalization and using the results from the following two subsections, one finds that the free energy (\ref{Ompoint})
is smaller than the free energy of the mesonic and chirally restored phases as soon as baryons are allowed to appear. Therefore, the chemical potential (\ref{muc})
marks the transition from vacuum to baryonic matter.

\subsubsection{Confined geometry}

The Lagrangian for pointlike baryons with maximally separated flavor branes in the confined geometry is 
\be
{\cal L} = {\cal L}_0+n_I\left[\frac{u_{\rm KK}}{3}-\hat{a}_0(u_{\rm KK})\right]\delta(u-u_{\rm KK}) \, , 
\ee
with
\be \label{L0Conf}
{\cal L}_0 = \frac{u^{5/2}}{\sqrt{f}}\sqrt{1-f \hat{a}_0'^2} \, .
\ee
The equation of motion for $\hat{a}_0$ in integrated form is thus
\be
\frac{u^{5/2}\sqrt{f}\hat{a}_0'}{\sqrt{1-f \hat{a}_0'^2}} = n_I \, .
\ee
By defining the new variable
\be
y(u) = \int_{u_{\rm KK}}^u \frac{dv}{\sqrt{f}\sqrt{v^5+n_I^2}} \, ,
\ee
the equation of motion becomes
\be
\partial_y \hat{a}_0 = n_I \quad \Rightarrow \qquad \hat{a}_0(y) = n_I y + c \, . 
\ee
We determine the integration constant $c$ with the help of the boundary condition $\hat{a}_0(y_\infty) = \mu$, where
\be
y_\infty \equiv y(u=\infty) = \int_{u_{\rm KK}}^\infty \frac{du}{\sqrt{f}\sqrt{u^5+n_I^2}} \, .
\ee
We find $c=\mu-n_Iy_\infty$ and thus
\be
\hat{a}_0(y) = n_I(y-y_\infty) + \mu \, .
\ee
Minimization with respect to $n_I$ yields [completely analogously to the deconfined geometry, see Eq.\ (\ref{a0uc})]
\be
\hat{a}_0(u_{\rm KK}) = \frac{u_{\rm KK}}{3} \, .
\ee
Consequently, since $y(u=u_{\rm KK})=0$, we obtain the following implicit equation for $n_I$,
\be
\mu-n_I y_\infty = \frac{u_{\rm KK}}{3} \, .
\ee
As in the deconfined geometry, the free energy only receives a contribution from ${\cal L}_0$. With
\be
\hat{a}_0' = \frac{dy}{du}\partial_y\hat{a}_0 = n_I \frac{dy}{du} \quad \Rightarrow \qquad 1-f\hat{a}_0'^2 = \frac{u^5}{u^5+n_I^2} \, , 
\ee
we obtain 
\be
\Omega = {\cal N}\int_{u_{\rm KK}}^\infty du\frac{u^5}{\sqrt{f}\sqrt{u^5+n_I^2}} \, .
\ee

\subsection{Mesonic Phase}
\label{sec:mesonic}

\subsubsection{Deconfined geometry}

For the mesonic phase, we simply set $n_I=0$ in the Lagrangian (\ref{Lpoint}), such that ${\cal L}={\cal L}_0$. 
This leads to the equations of motion in integrated form
\begin{subequations}
\bea
\frac{u^{5/2}\hat{a}_0'}{\sqrt{1+u^3f_Tx_4'^2-\hat{a}_0'^2}} &=& 0 \, , \\[2ex]
\frac{u^{5/2}u^3f_Tx_4'}{\sqrt{1+u^3f_Tx_4'^2-\hat{a}_0'^2}} &=& f_T^{1/2}(u_0)u_0^4 \label{eom2mes}\, , 
\eea
\end{subequations}
where we have denoted the location of the tip of the connected branes by $u_0$ (at the second-order onset of pointlike baryons, $u_0=u_c$), 
and where we have determined the integration constants 
on the right-hand side from the boundary conditions
\be \label{boundarymesonic}
\hat{a}_0'(u_0) = 0 \, , \qquad \hat{a}_0(\infty) = \mu \, , \qquad x_4'(u_0)=\infty \, , \qquad \frac{\ell}{2} = \int_{u_0}^\infty du\,x_4' \, .
\ee
Consequently,
\be
\hat{a}_0' = 0 \, , \qquad x_4' = \frac{f_T^{1/2}(u_0)u_0^4}{f_T^{1/2}(u) u^{3/2}\sqrt{u^8f_T(u)-u_0^8f_T(u_0)}} \, .
\ee
In particular, $\hat{a}_0$ is constant, indicating that the baryon density vanishes in this phase. It thus remains to determine $u_0$ as a function of temperature (it does not 
depend on $\mu$). This is done with the help of the last relation of Eq.\ (\ref{boundarymesonic}). 
At zero temperature, where $f_T(u)=1$, one finds the analytical solution
\be
u_0(T=0) = \frac{16\pi}{\ell^2}\left[\frac{\Gamma\left(\frac{9}{16}\right)}{\Gamma\left(\frac{1}{16}\right)}\right]^2 \simeq \frac{0.52486}{\ell^2} \, .
\ee
The free energy becomes
\be
\Omega_{\rm mesonic} = {\cal N}\int_{u_0}^\infty du\,u^{5/2}\frac{u^4 f_T^{1/2}(u)}{\sqrt{u^8f_T(u)-u_0^8f_T(u_0)}} \, .
\ee
Again, at $T=0$, this can be evaluated analytically. Introducing a cutoff $\Lambda$, we obtain
\be \label{OmMesT0}
\frac{\Omega_{\rm mesonic}(T=0)}{\cal N} 
= \int_{u_0}^\Lambda du\,u^{5/2}\frac{u^4}{\sqrt{u^8-u_0^8}} = \frac{2}{7}\Lambda^{7/2}-\underbrace{\frac{2^{15}\pi^4}{15\,\ell^7}\frac{\Gamma\left(\frac{31}{16}
\right)\tan\frac{\pi}{16}}{\Gamma\left(\frac{23}{16}\right)}\left[\frac{\Gamma\left(\frac{9}{16}\right)}{\Gamma\left(\frac{1}{16}\right)}\right]^7}_{\displaystyle{\simeq 5.4\times 10^{-3}/\ell^7}} \, .
\ee 
This value is used for example in the upper right panel of Fig.\ \ref{figdeconf2}.

\subsubsection{Confined geometry}

Setting $n_I=0$, the Lagrangian is simply given by ${\cal L}={\cal L}_0$ with ${\cal L}_0$ from Eq.\ (\ref{L0Conf}), which 
yields the equation of motion for $\hat{a}_0$ in integrated form 
\be
\frac{u^{5/2}\sqrt{f}\hat{a}_0'}{\sqrt{1-f \hat{a}_0'^2}} = 0 \, .
\ee
Here we have set the integration constant to zero to ensure $\hat{a}_0'(u_{\rm KK})=0$. As a consequence, the solution is trivial, $\hat{a}_0(u) = \mu$, 
and the free energy becomes
\be
\Omega = {\cal N}\int_{u_{\rm KK}}^\infty du\frac{u^{5/2}}{\sqrt{f}} \, .
\ee

\subsection{Chirally restored phase}
\label{sec:restored}

In the chirally restored phase, we start from the same Lagrangian as in the mesonic phase, but since the flavor branes are straight and disconnected, the holographic 
coordinate assumes values in the interval $u\in [u_T,\infty]$, and we have $x_4'=0$. Therefore, there is only one nontrivial equation of motion whose integration constant is the 
baryon density $n_I$,
\be
\frac{u^{5/2} \hat{a}_0'}{\sqrt{1-\hat{a}_0'^2}} = n_I \, .
\ee
In this phase, there are no baryons, the baryon density is generated by quarks.  With the boundary conditions
\be
\hat{a}_0(u_T) = 0 \, , \qquad \hat{a}_0(\infty) = \mu \, , 
\ee
we find 
\be
\hat{a}_0(u) = \mu-\frac{n_I^{2/5}\Gamma\left(\frac{3}{10}\right)\Gamma\left(\frac{6}{5}\right)}{\sqrt{\pi}}
+ u\,{}_2 F_1\left[\frac{1}{5},\frac{1}{2},\frac{6}{5},-\frac{u^5}{n_I^2}\right] \, , 
\ee
and the following implicit equation for $n_I$ as a function of $\mu$ and $T$,
\be \label{ceq}
0 = \mu-\frac{n_I^{2/5}\Gamma\left(\frac{3}{10}\right)\Gamma\left(\frac{6}{5}\right)}{\sqrt{\pi}}
+ u_T\,{}_2 F_1\left[\frac{1}{5},\frac{1}{2},\frac{6}{5},-\frac{u_T^5}{n_I^2}\right] \, .
\ee
At zero temperature, we have the simple result 
\be
n_I(\mu,T=0) = \frac{\pi^{5/4}\mu^{5/2}}{\left[\Gamma\left(\frac{3}{10}\right)\Gamma\left(\frac{6}{5}\right)\right]^{5/2}} \, .
\ee  
The free energy is
\be
\frac{\Omega_{\rm quark}}{\cal N} = \int_{u_T}^\infty du\,\frac{u^5}{\sqrt{u^5+n_I^2}} = \frac{2}{7}\Lambda^{7/2}-\frac{2\Gamma\left(\frac{3}{10}
\right)\Gamma\left(\frac{6}{5}\right)}{7\sqrt{\pi}}\,n_I^{7/5}-\frac{2u_T n_I}{7}\eta\left(\frac{u_T^{5/2}}{n_I}\right) \, ,
\ee
where again we have introduced a cutoff for $u\to \infty$, and abbreviated 
\be
\eta(x)\equiv \sqrt{1+x^2}-{}_2F_1\left[\frac{1}{5},\frac{1}{2},\frac{6}{5},-x^2\right] \, .
\ee
At zero temperature,
\be
\frac{\Omega_{\rm quark}(T=0)}{\cal N} = \frac{2}{7}\Lambda^{7/2} - \frac{2\pi^{5/4}\mu^{7/2}}{7\left[\Gamma\left(\frac{3}{10}\right)\Gamma\left(\frac{6}{5}\right)\right]^{5/2}} \, .
\ee
By comparing this free energy to the one from the mesonic phase (\ref{OmMesT0}), we compute the critical chemical potential for the chiral phase transition at $T=0$,
\be
\mu_c(T=0) = \frac{16\pi^{11/14}}{\ell^2}\left[\frac{7\Gamma\left(\frac{31}{16}\right)\tan\frac{\pi}{16}}{15\Gamma\left(\frac{23}{16}\right)}\right]^{2/7}
\left[\frac{\Gamma\left(\frac{9}{16}\right)}{\Gamma\left(\frac{1}{16}\right)}\right]^{2}\left[\Gamma\left(\frac{3}{10}\right)\Gamma\left(\frac{6}{5}\right)\right]^{5/7}
\simeq \frac{0.440472}{\ell^2} \, .
\ee
For nonzero temperatures, the critical chemical potential has to be computed numerically. The result is shown in Fig.\ \ref{figpointlike}, where we see that for small 
temperatures there is no chiral phase transition because the mesonic phase is superseded by the baryonic phase {\it before} the chirally restored phase becomes favored over the 
mesonic phase.

\bibliography{references}

\end{document}